\documentclass[12pt]{article}
\usepackage{fancyhdr}
\usepackage{latexsym} 
\usepackage{amssymb}
\usepackage{graphicx}
\usepackage{epsf}
\usepackage{amscd}
\usepackage{amsmath}

\def\ee{\end{eqnarray}}

\newcommand{\nn}{\nonumber}

\def\=:{=\hspace{-.7em}\raisebox{1.1ex}{.}\hspace{.1em}\raisebox{-0.2ex}{.} }

\makeatletter
\@addtoreset{equation}{section}

\renewcommand{\thefootnote}{\fnsymbol{footnote}}
\makeatother

\begin{document}
\thispagestyle{empty}

\begin{flushright}
TIT/HEP--543 \\
{\tt hep-th/0602289} \\
February, 2006 \\
\end{flushright}
\vspace{3mm}

\begin{center}
{\Large \bf 
Manifestly Supersymmetric Effective Lagrangians \\
on BPS Solitons 
} 
\\[12mm]
\vspace{5mm}

\normalsize
  {\large \bf 
Minoru~Eto}
\footnote{\it  e-mail address: 
meto@th.phys.titech.ac.jp
}, 
  {\large \bf 
Youichi~Isozumi}
\footnote{\it  e-mail address: 
isozumi@th.phys.titech.ac.jp
}, 
  {\large \bf 
Muneto~Nitta}
\footnote{\it  e-mail address: 
nitta@th.phys.titech.ac.jp
}, 
\\  {\large \bf 
 Keisuke~Ohashi 
}\footnote{\it  e-mail address: 
keisuke@th.phys.titech.ac.jp
}, 
~and~~  {\large \bf 
Norisuke~Sakai}
\footnote{\it  e-mail address: 
nsakai@th.phys.titech.ac.jp
} 

\vskip 1.5em

{ \it Department of Physics, Tokyo Institute of 
Technology \\
Tokyo 152-8551, JAPAN  
 }
\vspace{15mm}
{\bf Abstract}\\[5mm]
{\parbox{13cm}{\hspace{5mm}
A systematic method to obtain the effective Lagrangian on 
the BPS background in supersymmetric gauge theories is 
worked out, taking domain walls and vortices as concrete 
examples. 
The Lagrangian in terms of the superfields with four 
preserved supercharges is expanded in powers of the 
slow-movement parameter $\lambda$. 
The expansion gives the superfield form of the BPS 
equations at ${\cal O}(\lambda^0)$, and all the fluctuation 
fields at ${\cal O}(\lambda^1)$. 
The density of the K\"ahler potential for the effective 
Lagrangian follows as an automatic consequence of the 
$\lambda$ expansion making (four preserved) supercharges manifest. 

}}
\end{center}
\vfill
\newpage
\setcounter{page}{1}
\setcounter{footnote}{0}
\renewcommand{\thefootnote}{\arabic{footnote}}
\section{Introduction}\label{INTRO}
Various topological solitons have been considered in 
constructing models in the brane-world scenario 
\cite{HoravaWitten}--\cite{RandallSundrum}. 
It has been quite useful to consider supersymmetric 
gauge theories for phenomenological purposes \cite{DGSW}. 
The simplest solitons 
are domain walls giving one extra dimension, 
which have been studied 
extensively in theories with 
four supercharges~\cite{Cvetic:1991vp}--\cite{N1SUGRAwall} 
and in theories 
with eight supercharges \cite{N=2walls}--\cite{N2SUGRAwall}.
The next simplest solitons are 
vortices \cite{Abrikosov:1956sx}--\cite{Auzzi:2005yw} 
giving two extra dimensions.
Monopoles and Yang-Mills instantons are solitons 
with three and four extra dimensions, respectively. 
If a field configuration preserves a part of supersymmetry (SUSY), 
it satisfies the field equation automatically 
\cite{WittenOlive}. 
Such configuration is called the 
Bogomol'nyi-Prasad-Sommerfield 
(BPS) state \cite{BPS}. 
We usually obtain a family of BPS solutions characterized by 
parameters, called moduli. 
These moduli parameters constitute the moduli spaces.
Although solutions and their moduli spaces were established 
and discussed extensively for 
instantons \cite{Atiyah:1978ri,Corrigan:1983sv} 
and monopoles \cite{Nahm:1979yw,Corrigan:1983sv}, 
those for domain walls and vortices have been much less 
studied for a long time. 
Recently the moduli space of BPS domain walls in a supersymmetric 
non-Abelian gauge theory with eight supercharges 
has been completely characterized and explicit solutions 
have been found for strong gauge coupling with complete 
moduli and for finite coupling with partial moduli \cite{INOS1,INOS2}. 
This is achieved by solving the hypermultiplet BPS 
equation and by rewriting 
the remaining BPS equation into a 
``master equation'' for a gauge invariant quantity 
$\Omega$. 
The moduli space of vortices is also characterized in 
Abelian gauge theory \cite{Taubes:1979tm}--\cite{Chen:2004xu}. 
Vortices in non-Abelian gauge theory, 
called non-Abelian vortices, have been recently 
found \cite{Hanany:2003hp}--\cite{Eto:2006mz}, 
and their moduli space has been determined 
\cite{Hanany:2003hp,Eto:2005yh}. 
These solitons in the Higgs phase are extensively reviewed 
recently \cite{Eto:2006pg}. 
It is important to construct the low-energy effective 
Lagraigian of the localized modes on such solitons for 
brane-world scenario. 
In order to obtain the low-energy effective Lagrangian, 
the standard method is to promote the moduli parameters of 
the background soliton into fields on the world volume of 
the soliton \cite{Manton:1981mp}. 
The method is based on the assumption of the weak dependence 
on the world-volume coordinates, and gives the low-energy 
effective Lagrangian which contains all nonlinear terms 
with two derivatives or less. 
Another interesting aspect of the soliton dynamics is the 
scattering of solitons 
\cite{Manton:1981mp,MantonSutcliffe}: 
it has been extensively studied primarily 
for cases without the spacial world volume. 

The purpose of our paper is to establish a systematic 
method to obtain the effective Lagrangian on BPS background 
in supersymmetric gauge theories maintaining the preserved 
SUSY manifest. 
We explicitly work out our method by taking 
domain walls and vortices in the supersymmetric $U(N_{\rm C})$ 
gauge theories with eight supercharges  
with $N_{\rm F} (\geq N_{\rm C})$ hypermultiplets in the 
fundamental representation as illustrative examples. 
Although we work in the space-time dimensions highest 
allowed by supersymmetry, namely vortices and walls in 
six and five dimensions, respectively, 
our discussion is applicable completely in lower dimensions 
which can be obtained simply by dimensional reductions. 
Since we can naturally specify the order of magnitude in 
powers of the slow-movement parameter $\lambda$ for 
various fields, we obtain a systematic expansion of the 
Lagrangian in powers of $\lambda$. 
Then a superfield form of the BPS equations results at the 
zero-th order in $\lambda$, and the superfield equation to 
determine all the fluctuation fields follows 
at the next order. 
Here we have retained up to the terms of order $\lambda^2$ 
in the Lagrangian, 
in order to obtain the effective Lagrangian 
at the lowest nontrivial order, namely up to two 
derivatives. 
We anticipate that retaining higher powers of 
$\lambda$ in our systematic expansion will offer a 
systematic method to compute the effective Lagrangian 
with higher derivative terms. 
Since four SUSY are preserved manifest throughout 
our procedure, our result is summarized 
as a density of the K\"ahler potential in four SUSY 
superspace. 
By integrating over the extra dimensions, the K\"ahler 
potential of the effective Lagrangian is obtained. 
Our results should be useful to study soliton scattering 
in $U(N_{\rm C})$ gauge theories. 
We have also worked out explicitly the effective 
Lagrangian for multi-wall systems. 
For finite gauge coupling, we have illustrated the use of 
our effective Lagrangian by solving the $1/2$ BPS lumps 
in the double-wall effective Lagrangian and found that 
the energy of boojum \cite{INOS3}, \cite{Sakai:2005sp} 
is correctly reproduced.

A number of formulations of theories with eight supercharges 
in space-time dimensions greater than four 
have been devised to use superfields 
maintaining only the four supercharges manifest 
\cite{MSS}--\cite{KakimotoSakai}. 
We have succeeded to obtain 
a manifestly supersymmetric method to determine the low-energy 
effective Lagrangian using the superfields for 
four preserved SUSY. 
Since we are interested in topological solitons whose 
energies are generally given by topological charges, it is 
necessary to keep track of all the total derivative terms, 
which are often neglected in using the superfield 
formalism for four SUSY. 
Moreover, it is important to realize that the auxiliary 
fields of the superfields 
for the four 
SUSY is different from auxiliary fields for the eight SUSY 
by total derivative terms. 
From these two facts, we find that 
the topological charges follow as an automatic consequence 
of rewriting 
the fundamental (five or six dimensional) 
Lagrangian in terms of the superfield for four SUSY, 
in the cases of domain walls or vortices, respectively.
Namely, the Lagrangian in terms of the superfield with 
four preserved SUSY manifest is different from the 
fundamental Lagrangian with 
manifest eight SUSY by a total 
divergence term which precisely 
gives the topological charge of the BPS soliton. 

To obtain the K\"ahler potential of the effective Lagrangian, 
we first solve the BPS equation for 
hypermultiplets and rewrite the Lagrangian in terms of 
the remaining dynamical degree of freedom, the gauge invariant 
quantity $\Omega$. 
The result is given by the K\"ahler potential 
density $K(\Omega, y)$ in terms of the gauge invariant quantity 
$\Omega$ as the dynamical variable. 
If we replace $\Omega$ by the solution $\Omega_{\rm sol}$ 
of the master equation 
and integrate over the extra dimensions, 
we finally obtain the effective Lagrangian. 
One interesting feature 
is that the K\"ahler potential density $K(\Omega, y)$ 
in five or six dimensions (before $\Omega$ is replaced 
by $\Omega_{\rm sol}$) 
serves as a Lagrangian from which the 
master equation for the gauge invariant quantity $\Omega$ 
can be derived by the usual minimal action principle. 
Our derivation of the K\"ahler potential density 
explains this empirical observation, 
since the Lagrangian with the K\"ahler potential density 
can be understood as the fundamental Lagrangian 
after the functional integral over 
the hypermultiplet in this approximation.

In sect.\ref{sc:slow-move}, 
after a brief review of 
component formalism for walls, 
we express the fundamental 
Lagrangian in terms of superfields with four SUSY and 
expand the Lagrangian in five dimensions in powers of 
the slow-movement parameter $\lambda$, to 
obtain the density of the K\"ahler potential for the 
effective Lagrangian on $1/2$ BPS walls. 
In sect.\ref{sc:vortex}, a superfield treatment of the 
effective Lagrangian on vortices is worked out as another 
example to show the usefulness of our method. 
In sect.\ref{sc:eff-Lag-multi-wall}, 
we obtain explicitly the effective Lagrangian of multi-walls 
for exact solutions at strong coupling limit as well as 
at a discrete finite value of gauge coupling. 
Using double-wall effective Lagrangian, lumps are worked 
out to give boojum correctly. 
Sect.\ref{sc:discussion} is devoted to brief discussion. 
Appendix \ref{sc:component-approach} contains a 
component approach for the wall case. 
Appendix \ref{sc:variation-6d-Lag} contains useful formulas 
in the six-dimensional Lagrangian for the vortex case.

\section{Slow-move Approximation for Walls
}
\label{sc:slow-move}

\subsection{Component Formalism of Slow-move Approximation}
\label{sc:slow-move-comp}

Let us here review briefly our model and the usual 
component method to solve 
the BPS equations. 
The bosonic parts of the Lagrangian with a 
common gauge coupling constant $g$ for $U(N_{\rm C})$  
in five dimensions is given by 
\begin{eqnarray}
&\!\!\!&\!\!\!{\cal L}|_{\rm boson} 
= 
{\rm Tr}\biggl[-{1\over 2g^2} F_{MN}(W)F^{MN}(W)
-\frac{1}{g^2}({\cal D}_M \Sigma)^2 
+
\frac{1}{g^2}(Y^a)^2-c^aY^a 
\nonumber \\
&\!\!\!-&\!\!\!  {\cal D}^M H^i ({\cal D}_M H^i)^\dagger 
+F^iF^{i\dagger} 
-(\Sigma H^i-H^iM)(\Sigma H^i-H^iM)^\dagger 
+Y^a(\sigma^a)_{ij}H^j H^{i\dagger} 
\biggr] .
\label{fundamental-Lag2}
\end{eqnarray}
Here the bosonic components in the vector multiplet 
are gauge fields $W_M$, the real scalar fields $\Sigma$ 
and the triplet of auxiliary fields $Y^a, a=1,2,3$, all 
in the adjoint representation, 
and those in the hypermultiplet are the doublets of 
the complex scalar fields $H^i$ and the auxiliary fields 
$F^i$, $i=1,2$ which can be assembled into 
$N_{\rm C}\times N_{\rm F}$ matrices. 
The indices $M, N=0, 1,\cdots, 4$ run over five-dimensions, 
and the mostly plus signature is used for the 
metric $\eta_{MN}={\rm diag}.(-1, +1, \cdots, +1)$. 
The covariant derivatives are defined as 
$D_M \Sigma = \partial_M \Sigma + i[ W_M , \Sigma ]$, 
$D_M H^{i}=(\partial_M + iW_M)H^{i} 
$, 
and field strength is defined as 
$F_{MN}=\frac{1}{i}[D_M , D_N]
=\partial_M W_N -\partial_N W_M + i[W_M, W_N]$. 
After eliminating auxiliary fields $Y^a$, 
the scalar potential $V$ is given by 
\begin{eqnarray}
V&\!\!\!=&\!\!\! 
\frac{g^2}{4}
{\rm Tr}
\Big[
\left(
H^{1}  H^{1\dagger}  - H^{2} H^{2\dagger} 
- c\mathbf{1}_{N_{\rm C}}
\right)^2 +
 4 H^2H^{1\dagger} H^1H^{2\dagger}
\Big] 
\nonumber\\
&\!\!\!
&\!\!\!
+{\rm Tr}\left[
 (\Sigma H^i - H^i M) 
 (\Sigma H^i - H^i M)^\dagger 
 \right],
\end{eqnarray}
with the hypermultiplet 
mass matrix $M={\rm diag}(m_1,\,\cdots, m_{N_{\rm F}})$ 
and the Fayet-Iliopoulos parameter 
taken along the 
third direction in $SU(2)_R$ as 
$c_a =(0,\ 0,\ c)$ with $c>0$. 

By requiring half of SUSY to be preserved, we obtain 
the $1/2$ BPS equations for domain walls which depend on 
$y$ only 
\begin{eqnarray}
{\cal D}_y H^1 &\!\!\!=&\!\!\! -\Sigma H^1 + H^1 M,\qquad 
{\cal D}_y H^2 = \Sigma H^2 -H^2 M,\label{BPSeq-H}
\\
{\cal D}_y \Sigma &\!\!\!=&\!\!\! 
{g^2\over 2}\left(c{\bf 1}_{N_{\rm C}}-H^1H^1{}^\dagger 
+H^2H^2{}^\dagger \right), 
\label{BPSeq-Sigma}
\qquad 
0 = 
g^2 H^1H^2{}^\dagger . 
\end{eqnarray}
The solution of the BPS equations saturates the BPS bound for 
the tension of the (multi-)wall  
\begin{eqnarray}
T_{\rm w}&\!\!=\!\!&\!\!
\int^{+\infty}_{-\infty}\hspace{-1.5em}dy{\cal E}_{\rm w}
=\int^{+\infty}_{-\infty}\hspace{-1.5em}dy
\partial_y\Big[{\rm Tr}\big[c\Sigma 
-(\Sigma H^1H^{1\dagger }-H^1MH^{1\dagger })
+(\Sigma H^2H^{2\dagger }-H^2MH^{2\dagger })\big]\Big]
\nn\\
&\!\!\!\!=&\!\!\!\!
c \left[{\rm Tr}\Sigma \right]^{+\infty}_{-\infty}
\label{eq:tension}
\end{eqnarray}
where the energy density is denoted as ${\cal E}_{\rm w}$. 
It has been shown that the hypermultiplet BPS equation 
(\ref{BPSeq-H}) can be solved by \cite{INOS1}, \cite{INOS2} 
\begin{eqnarray}
 \Sigma +iW_y&\!\!\!=&\!\!\!S^{-1}(y)\partial _yS(y), 
\qquad 
W_\mu =0, \qquad 
(\mu =0,\cdots,3) 
\label{eq:comp-gauge-tr} 
\\
H^1&\!\!\!=&\!\!\! S^{-1}(y)H_0e^{My}, \qquad H^2=0, 
\label{eq:hyper-sol} 
\end{eqnarray}
where the moduli matrix $H_0$ is obtained as an integration 
constant carrying all the parameters of the solution, namely 
moduli. 
The moduli matrices related by the following $V$-equivalence 
transformations are physically equivalent: 
\begin{eqnarray}
 H_0 \rightarrow V H_0, 
\qquad 
S(y) \rightarrow V S(y), 
\qquad 
V \in GL(N_{\rm C}, {\bf C}). 
\label{eq:v_equivalence} 
\end{eqnarray}
The vector multiplet BPS equation (\ref{BPSeq-Sigma}) 
can be converted to the following 
``master equation'' for a gauge invariant 
quantity $\Omega \equiv SS^\dagger $ \cite{INOS1} 
\begin{eqnarray}
 \partial _y\left(\Omega ^{-1}\partial _y\Omega \right)
=g^2c\left({\bf 1}_{N_{\rm C}}-\Omega ^{-1}\Omega _0\right),
\qquad \Omega _0\equiv c^{-1}H_0e^{2My}H_0^\dagger. 
\label{eq:master-eq}
\end{eqnarray}
The matrix function $S$ can be determined from the solution 
$\Omega$ of this master equation by fixing a gauge, 
and all the other fields can be obtained from $S$ and $H_0$. 
Since the BPS soliton has co-dimension one, the solution 
represents (multiple) domain walls. 
There are two characteristic mass scales in this system: 
mass differences $\Delta m$ of hypermultiplets, and 
the mass scale in front of the master equation $g\sqrt{c}$. 
In the strong coupling limit $g\sqrt{c}\gg \Delta m$, 
the vector multiplet serves 
to give constraints to hypermultiplets leading 
to the 
nonlinear sigma model~\cite{ANS}, whose BPS domain wall 
solutions have 
been obtained exactly \cite{INOS1}, \cite{INOS2}. 

The low-energy effective Lagrangian on the world volume of 
solitons is given by promoting the moduli parameters to 
fields on the soliton and by assuming the weak dependence 
on the world-volume coordinates of the soliton 
\cite{Manton:1981mp}. 
In the case of domain walls, all the moduli parameters 
are contained in the moduli matrix $H_0$ and 
constitute the complex Grassmann manifold \cite{INOS1}, 
\cite{INOS2}. 
Denoting the inhomogeneous coordinates of the complex 
Grassmann manifold in the moduli matrix $H_0$ 
as $\phi ^\alpha $, we promote them to 
fields which depend on $x^\mu , (\mu =0,1,2,3)$ 
\begin{eqnarray}
 H_0(\phi ^\alpha )\rightarrow H_0(\phi ^\alpha (x)) , 
\label{eq:moduli-fields}
\end{eqnarray}
which has also been studied in our recent review article 
\cite{Eto:2006pg}. 
We introduce ``the slow-movement parameter'' $\lambda$, 
which is assumed to be much smaller than the typical 
mass scale in the problem, in our case, 
$\Delta m$ and $g\sqrt{c}$.  
\begin{eqnarray}
 \lambda \ll {\rm min}(\Delta m, g\sqrt{c} ).
\end{eqnarray}
The nonvanishing fields of the $1/2$ BPS background have 
contributions independent of $\lambda$, and 
derivatives in terms of the world volume coordinates 
are assumed to be of order $\lambda$, expressing 
the weak dependence on the world-volume coordinates 
\begin{eqnarray}
H^1 \sim {\cal O}(1),\quad \Sigma \sim {\cal O}(1),\quad 
\partial _\mu \sim {\cal O}(\lambda ). 
\label{eq:h1-order}
\end{eqnarray}
Those fields which vanish in the background solution can 
now have nonvanishing values, induced by the fluctuations 
of the moduli fields of order $\lambda$ 
\begin{eqnarray}
W_\mu \sim {\cal O}(\lambda ),\quad 
H^2\sim {\cal O}(\lambda ) ,
\label{eq:h2-order}
\end{eqnarray}
\begin{eqnarray}
  {\cal D}_\mu H^1\sim {\cal O}(\lambda ),
\quad {\cal D}_\mu \Sigma\sim {\cal O}(\lambda ),
\quad F_{\mu y}(W)\sim {\cal O}(\lambda ), 
\label{eq:field-st-order}
\end{eqnarray}
and other components of the field strength are higher 
orders in $\lambda$. 
If we decompose the field equations in powers of 
$\lambda$, we find that order $\lambda ^0$ equations 
are automatically satisfied by the BPS equations 
(\ref{BPSeq-H}) and (\ref{BPSeq-Sigma}). 
However, it becomes more and more complicated to 
solve the field equation at higher orders in the 
expansion in powers of $\lambda$, since various fields 
that vanish in the background become nonvanishing, and 
need to be solved. 
For instance the equation of motion for the gauge field 
fluctuations $W_\mu$ reads 
\begin{eqnarray}
0&=&{1\over g^2}{\cal D}_yF_{\mu y}
+{i\over g^2}[\Sigma ,\,{\cal D}_\mu \Sigma ]
+{i\over 2}\left(H^1 {\cal D}_\mu H^{1\dagger} 
-{\cal D}_\mu H^1 \,H^{1\dagger} \right) 
, 
\label{eq:EOM-gauge-field}
\end{eqnarray}
as given in Appendix \ref{sc:component-approach}. 
To obtain the solution of this equation, one needs to 
do a long and tedious calculation, leading finally to 
Eq.(\ref{eq:sol-gauge-field}) in Appendix 
\ref{sc:component-approach}. 
Further long calculations give the K\"ahler 
metric and K\"ahler potential for the effective Lagrangian 
in Eqs.(\ref{eq:kahler-metric-wall}) and 
(\ref{eq:kahler-potential-wall}) 
in Appendix \ref{sc:component-approach}. 
The basic reason for these complications is that component 
fields are used to expand in powers of $\lambda$ 
without exploiting the constraint of SUSY. 
We leave a brief outline of this procedure in terms of 
component fields to Appendix \ref{sc:component-approach}, 
in order to facilitate a comparison 
to our result in terms of superfields. 
We shall show in the next section that maintaining the 
preserved SUSY manifest greatly helps to determine these 
newly nonvanishing fields and to organize the expansion of 
field equations in powers of $\lambda$.

\subsection{Superfield Formalism of Slow-move Approximation
}

The BPS wall background conserves a half of SUSY.  
Thus an action for fluctuations around the BPS background 
can be written in term of superfield respecting 
the surviving half of SUSY. 

Let us define the superfields\footnote{ 
We use the convention of Wess and Bagger \cite{WessBagger} 
for Grassmann coordinates and superfields in this paper, 
except that four-dimensional 
spacetime indices are denoted by Greek alphabets 
$\mu, \nu=0, \cdots, 3$. 
For conventions of superfields in terms of component fields, 
we mostly follow those in Refs.\cite{Hebecker}, and 
\cite{KakimotoSakai}. } 
using two component spinor Grassmann 
coordinates
$\theta^\alpha, \theta_{\dot \alpha}$. 
The components of superfields are fields in five dimensions. 
A vector multiplet with eight SUSY consists of 
a real vector superfield ${\bf V} (={\bf V}^\dagger)$ 
and an adjoint chiral superfield ${\bf \Phi}$ 
($\bar D_{\dot\alpha} {\bf \Phi} = 0$) 
in terms of superfield with four superchages. 
The vector superfield ${\bf V}$ 
contains a gauge field $W_\mu, \mu=0, \cdots, 3$ 
for the four spacetime dimensions, the half of gaugino 
field $\lambda_+$, and an auxiliary field ${\cal Y}^3$. 
If one takes the Wess-Zumino gauge, it becomes explicitly as 
\begin{equation}
{\bf V}\Big|_{\rm WZ}=
-\theta \sigma^\mu \bar\theta W_\mu 
+ i\theta^2\bar \theta \bar \lambda_+
- i\bar \theta^2 \theta \lambda_+ 
+{1 \over 2} \theta^2\bar \theta^2 {\cal Y}^3, 
\qquad {\cal Y}^3\equiv Y^3-{\cal D}_y\Sigma,  
\label{eq:vector-superfield}
\end{equation}
where the auxiliary field ${\cal Y}^3$ of the superfield 
for four SUSY is shifted from the auxiliary field $Y^3$ 
for eight SUSY by the covariant derivative of adjoint 
scalar $\Sigma$ along the fifth coordinate (the extra 
dimensions) $y$ \cite{Hebecker}, \cite{KakimotoSakai}. 
This difference becomes important in identifying the 
topological charge later. 
The chiral scalar superfield ${\bf \Phi}$ contains a 
complex scalar field made of the adjoint scalar $\Sigma$ 
and the fifth component of the gauge field $W_y$ as the 
real and imaginary part respectively, and the other half 
of gaugino $\lambda_-$ and a complex auxiliary field 
$Y^1+iY^2$ 
\begin{equation}
{\bf \Phi }
=\Sigma + i W_y + \sqrt{2}\theta (-i\sqrt{2} \lambda_-) 
+ \theta^2 (Y^1 +i Y^2). 
\label{eq:vectorr-chiral-superfield}
\end{equation}
The hypermultiplets are represented by a chiral superfields 
${\bf H}^1$ and an anti-chiral superfield ${\bf H}^2$. 
The (anti-) chiral superfield ${\bf H}^1$ (${\bf H}^2$) 
consists of the physical complex scalar field ${H}^1$ 
(${H}^2$), hyperino $\psi_+$ ($\psi_-$), and a complex 
auxiliary field ${\cal F}^1$ (${\cal F}^2$) 
\begin{eqnarray}
{\bf H}^1=H^1+\sqrt{2}\theta \psi_++\theta^2 {\cal F}^1, 
\qquad 
{\cal F}^1\equiv F^1+({\cal D}_y-\Sigma )H^2+H^2M, 
\label{eq:hyper-superfield1}
\end{eqnarray}
\begin{eqnarray}
{\bf H}^2=H^2+\sqrt{2}\bar \theta \bar \psi_-
+\bar \theta^2 {\cal F}^2, 
\qquad 
{\cal F}^2\equiv -F^2-({\cal D}_y+\Sigma )H^1+H^1M, 
\label{eq:hyper-superfield2}
\end{eqnarray}
where the auxiliary field ${\cal F}^1$ (${\cal F}^2$) 
of the superfield for four SUSY is shifted from 
the auxiliary field $F^1$ ($F^2$) 
for eight SUSY by the covariant derivative of the 
other hypermultiplet scalar $H^2$ ($H^1$) and 
other\footnote{
The other terms involving the adjoint scalar $\Sigma$ 
and the hypermultiplet mass matrix $M$ can be 
understood as a result of the Scherk-Schwarz dimensional 
reduction from six dimensions. 
} 
terms~\cite{Hebecker}, \cite{KakimotoSakai}. 
Please note that we have chosen to denote the anti-chiral 
superfield as ${\bf H}^2$, as shown in the $\bar \theta$ 
dependence in Eq.(\ref{eq:hyper-superfield2}). 

The complexified $U(N_{\rm C})$ gauge transformation is 
given in terms of the chiral scalar superfield 
${\bf \Lambda}$ for the gauge parameter by \cite{Hebecker}, 
\cite{KakimotoSakai}
\begin{eqnarray}
e^{2{\bf V}} \rightarrow e^{2{\bf V}'}
=e^{{\bf \Lambda}^\dagger} e^{2{\bf V}}e^{{\bf \Lambda}}, 
\quad 
{\bf \Phi } \rightarrow {\bf \Phi }'
=e^{-{\bf \Lambda}} {\bf \Phi }e^{{\bf \Lambda} }
+e^{-{\bf \Lambda}} \partial _y(e^{{\bf \Lambda} }),
\label{eq:gauge-tr-vector}
\end{eqnarray}
\begin{eqnarray}
{\bf H}^{1} \rightarrow {\bf H}^{1'}
=e^{-{\bf \Lambda}} {\bf H}^1,\quad 
{\bf H}^{2} \rightarrow {\bf H}^{2'}
=e^{{\bf \Lambda} ^\dagger }{\bf H}^2.
\label{eq:gauge-tr-hyper}
\end{eqnarray}
Then their infinitesimal transformations 
$\delta _{G}^{\bf C}({\bf \Lambda} )$ become 
\begin{eqnarray}
\delta _{G}^{\bf C}({\bf \Lambda} )e^{2{\bf V}}
&=&{\bf \Lambda}^\dagger e^{2{\bf V}}
+e^{2{\bf V}}{\bf \Lambda} ,\quad 
\delta _{G}^{\bf C}({\bf \Lambda} ){\bf \Phi }
=-[{\bf \Lambda} ,\,{\bf \Phi }]+\partial _y{\bf \Lambda} , 
\label{eq:infinitesimal-gauge1}
\\
\quad \delta _{G}^{\bf C}({\bf \Lambda} ){\bf H}^{1}
&=&-{\bf \Lambda} {\bf H}^1,
\quad \delta _{G}^{\bf C}({\bf \Lambda} ){\bf H}^{2}
={{\bf \Lambda}^\dagger }{\bf H}^2.
\label{eq:infinitesimal-gauge2}
\end{eqnarray}
Using the infinitesimal form of the gauge transformation, 
we can define the derivative $\hat D_y$ which is covariant 
under the complexified gauge transformations 
\begin{eqnarray}
{\hat D}_y\equiv \partial _y-\delta _{G}^{\bf C}({\bf \Phi }),  
\end{eqnarray}
For example, the covariant derivative for the hypermultiplet 
${\bf H}^1$ and the adjoint chiral scalar multiplet 
${\bf \Phi}$ are given by 
\begin{eqnarray}
{\hat D}_y{\bf H}^1
&\!\!\!=&\!\!\!(\partial _y+{\bf \Phi}){\bf H}^1, 
\nonumber \\ 
{\hat D}_ye^{2{\bf V}}&\!\!\!= &\!\!\!
\partial_y e^{2{\bf V}}-{\bf \Phi}^\dagger e^{2{\bf V}}-
e^{2{\bf V}}{\bf \Phi}. 
\end{eqnarray}

If supplemented by fermionic terms, the 
bosonic Lagrangian 
(\ref{fundamental-Lag2}) becomes invariant under the 
supersymmetric transformations with eight (real) 
Grassmann parameters. 
We can now rewrite this fundamental Lagrangian ${\cal L}$ 
in terms of the superfields for four supercharges as 
\begin{eqnarray}
 {\cal L}
&\!\!\!=&\!\!\!-{\cal E}_{\rm w} 
+ \int d ^4\theta{\rm Tr}\left[
-2c{\bf V}
+{1\over 2g^2}\left(
e^{-2{\bf V}}{\hat D}_ye^{2{\bf V}}\right)^2
+e^{2{\bf V}}{\bf H}^1{\bf H}^{1\dagger }
+e^{-2{\bf V}}{\bf H}^2{\bf H}^{2\dagger }
\right]\nn\\
&\!\!\!&\!\!\!{}+\left(\int d ^2\theta 
{\rm Tr}\left[
{\hat D}_y{\bf H}^1{\bf H}^{2\dagger }
-{\bf H}^1M{\bf H}^{2\dagger }
+{1\over 4g^2}{\bf W}^\alpha {\bf W}_\alpha 
\right]+{\rm h.c.}\right), 
\label{eq:fund-Lag-5d-4susy}
\end{eqnarray}
where field strength superfield ${\bf W}$ is given by 
\begin{eqnarray}
{\bf W}_\alpha \equiv -{1 \over 8} \bar D \bar 
D e^{-2{\bf V}} D_\alpha e^{2{\bf V}} . 
\end{eqnarray}
In transforming the fundamental Lagrangian 
(\ref{eq:fund-Lag-5d-4susy}) in terms of the 
superfield for four SUSY into the manifestly supersymmetric 
form for eight SUSY (\ref{fundamental-Lag2}), 
we need to make several partial 
integrations with respect to the fifth coordinate $y$, and 
have to retain the surface terms carefully 
in the procedure. 
We also note that the 
auxiliary fields for four SUSY ${\cal Y}^3$, 
and ${\cal F}^i$ are 
different from those for eight SUSY $Y^3$, $F^i$ 
by total derivative terms as in 
Eqs.(\ref{eq:vector-superfield}), 
(\ref{eq:hyper-superfield1}),  and 
(\ref{eq:hyper-superfield2}). 
In this way we find a total divergence ${\cal E}_{\rm w}$ 
representing the topological charge contributing to the 
energy density of the background which maintains four SUSY. 
Since we are interested in bosonic components of the 
toplogical term ${\cal E}$, we exhibit only the bosonic 
terms explicitly 
\begin{eqnarray}
{\cal E}_{\rm w}
&\!\!\!=&\!\!\!\partial_y\Big[{\rm Tr}\big[c\Sigma 
-(\Sigma H^1H^{1\dagger }-H^1MH^{1\dagger })
+(\Sigma H^2H^{2\dagger }-H^2MH^{2\dagger })\nn\\
&\!\!\!&\!\!\! {}
-{2\over g^2}{\cal Y}^3\Sigma 
+{\cal F}^1H^{2\dagger }+H^2{\cal F}^{1\dagger }
+({\rm fermionic~terms})\big]\Big]
.
\label{eq:total-der}
\end{eqnarray}
Let us emphasize again that the topological term is precisely 
the difference between the fundamental Lagrangian which 
is manifestly supersymmetric under the eight SUSY and 
another fundamental Lagrangian in terms of superfields 
for four manifest SUSY. 

Since we are interested in co-dimension one solitons, 
we are primaryly considering (multiple) domain walls 
as the $1/2$ BPS background. 
The solution typically have a number of parameters 
such as the position of walls. 
We call these parameters as moduli which are denoted 
as $\phi^i$. 
Following the usual procedure\cite{Manton:1981mp},  
we promote these moduli to fields $\phi^i(x)$ on 
the world volume of the background soliton, and 
assume that the moduli fields $\phi ^i(x)$ around the 
wall background to fluctuate only very slowly. 
Namely, we introduce a parameter $\lambda$ for the 
slow movement and neglect high energy fluctuations 
as explained in sect.\ref{sc:slow-move-comp}. 
By explicitly writing the derivatives of moduli fields 
we obtain 
\begin{eqnarray}
 \partial _y\phi ^i={\cal O}(1)\phi ^i,\qquad 
\partial _\mu \phi ^i={\cal O}(\lambda )\phi ^i,\quad  
\lambda \ll {\rm min}(\Delta m, g\sqrt{c}). 
\label{eq:slow-move1}
\end{eqnarray}
Here and in the following, ${\cal O}(1)$ means that it is 
of the order of the characteristic mass scale 
${\rm min}(\Delta m, g\sqrt{c})$. 
The slow-movement 
parameter $\lambda$ in Eq.(\ref{eq:slow-move1}) 
is defined to be of the order of the 
world-volume-coordinate derivative $\partial_\mu$. 
The supertransformation implies that the square of 
the derivative in terms of the Grassmann coordinates $\theta$ 
gives translation in the world-volume : 
$(\partial/\partial\theta)^2 \sim \partial_\mu$. 
Therefore we obtain 
\begin{eqnarray}
 d\theta \sim {\partial \over \partial \theta }
\sim {\cal O}(\lambda ^{1\over 2}). 
\end{eqnarray}
To assign the order of $\lambda$ for hypermultiplets, 
we observe that the first hypermultiplet $H^1$ has 
nonvanishing values whereas the second hypermultiplet $H^2$ 
vanishes in the $1/2$ BPS background solution (\ref{eq:hyper-sol}). 
If we let the moduli parameters to fluctuate over the 
world-volume coordinates with the order of $\lambda$, 
the fluctuation induces terms of order $\lambda$ in both 
hypermultiplets. 
Therefore the second hypermultiplet $H^2$ naturally 
becomes nonvanishing values and is of order $\lambda$. 
Combining the above order estimates of component fields, 
we assume that the hypermultiplet superfields 
are of order 
\begin{eqnarray}
{\bf H}^1\sim {\cal O}(1),\qquad 
{\bf H}^2\sim {\cal O}(\lambda ) .
\label{BPS-cond-by-order-assign-wall-1}
\end{eqnarray}
Note that this assignment breaks half of supersymmetry, 
and surviving supersymmetry is manifest in this 
superfield formalism. 
BPS equations for walls also respect 
this supersymmetry 
automatically, as we will explain later.   
Similarly, the adjoint scalar 
$\Sigma$ has nonvanishing values 
in the $1/2$ BPS background solution (\ref{eq:comp-gauge-tr}) 
\begin{eqnarray}
{\bf \Phi} \sim {\cal O}(1).
\label{BPS-cond-by-order-assign-wall-2}
\end{eqnarray}
On the other hand, the gauge field $W_\mu$ vanishes in 
the BPS background, and only induced by the order $\lambda$ 
fluctuations of moduli fields. 
Since the gauge field appears as the coeeficient of 
$\bar \theta \gamma^\mu \theta \sim {\cal O}(\lambda^{-1})$, 
we find the vector multiplet to be of the order of 
\begin{eqnarray}
{\bf V}\sim {\cal O}(1),\qquad 
(W_\mu \sim {\cal O}(\lambda )). 
\label{BPS-cond-by-order-assign-wall-3}
\end{eqnarray}
Then the kinetic terms for ${\bf V}, {\bf H}^2$ become of the order of 
\begin{eqnarray}
 \int d^2\theta {\rm Tr}W^\alpha W_\alpha 
\sim {\cal O}(\lambda ^4), 
\qquad \int d^4\theta {\rm Tr}
e^{-2{\bf V}}{\bf H}^2{\bf H}^{2\dagger }
\sim {\cal O}(\lambda ^4).
\end{eqnarray}
Since we are interested in obtaining the leading nontrivial 
terms in the expansion in powers of $\lambda$, we retain 
up to two powers of $\lambda$ corresponding to the nonlinear 
kinetic terms for the moduli fields. 
Neglecting ${\cal O}(\lambda ^4)$ 
we obtain 
\begin{eqnarray}
{\cal L}
&\!\!\!=&\!\!\! 
-{\cal E}_{\rm w}
+\int d^4\theta {\rm Tr}\left[
-2c {\bf V}
+e^{2 {\bf V}}{\bf H}^1{\bf H}^{1\dagger }
+{1\over 2g^2}\left(
e^{-2{\bf V}}{\hat D}_ye^{2{\bf V}}\right)^2
\right]\nn\\
&\!\!\!&\!\!\!{}+\left(\int d^2 \theta  
{\rm Tr}\left[{\hat D}_y{\bf H}^1{\bf H}^{2\dagger }
-{\bf H}^1M{\bf H}^{2\dagger }\right]+{\rm h.c.}\right). 
\label{eq:Lag-order2}
\end{eqnarray}

Up to this order, we can see that 
${\bf H}^2$, ${\bf V}$ serve as Lagrange multiplier fields. 
Namely the field equations for ${\bf H}^2$ and ${\bf V}$ 
give constraints 
\begin{eqnarray}
\hat D_y{\bf H}^1
&\!\!\!=&\!\!\!{\bf H}^1M, 
\label{eq:superfield-constr2}\\
g^2(c-{\bf H}^1{\bf H}^{1\dagger }e^{2{\bf V}}) 
&\!\!\!=&\!\!\!
- \hat D_y\left(e^{-2{\bf V}}\hat D_ye^{2{\bf V}}\right), 
\label{eq:superfield-constr1}
\end{eqnarray} 
respectively. 
These constraint equations are in terms of superfields. 
To clarify the physical content of the constraint 
equations (\ref{eq:superfield-constr2}) and 
(\ref{eq:superfield-constr1}), let us expand them 
in powers of the Grassmann coordinates 
$\theta, \bar \theta$. 
The first constraint equation 
(\ref{eq:superfield-constr2}) gives the 
BPS equation (\ref{BPSeq-H}) for the hypermultiplet $H^1$ 
as the lowest term (independent of $\theta, \bar \theta$). 
For the second constraint equation 
(\ref{eq:superfield-constr1}), 
it is useful to multiply $e^{\bf V}$ 
and $e^{-{\bf V}}$ from left and right. 
Then the left-hand side becomes 
\begin{eqnarray}
e^{{\bf V}} 
g^2(c-{\bf H}^1{\bf H}^{1\dagger }e^{2{\bf V}})
e^{-{\bf V}}
=
g^2\left((c- H^1 H^{1\dagger })
+
\theta\sigma^\mu\bar\theta
i(
{\cal D}_\mu H^1H^{1\dagger }
-H^1{\cal D}_\mu H^{1\dagger })
\right)+\cdots ,
\label{eq:constr-vector-comp2}
\end{eqnarray}
where dots denote terms with other combinations of 
$\theta$, and terms bilinear in fermions are neglected. 
The right-hand side of Eq.(\ref{eq:superfield-constr1}) 
becomes 
\begin{eqnarray}
-e^{{\bf V}}\hat D_y\left(
e^{-2{\bf V}}\hat D_ye^{2{\bf V}}\right)e^{-{\bf V}}
=
2\left({\cal D}_y \Sigma 
-\theta\sigma^\mu\bar\theta\left({\cal D}_yF_{\mu y}
+i[\Sigma ,{\cal D}_\mu \Sigma ]\right) 
\right)+\cdots. 
\label{eq:constr-vector-comp1}
\end{eqnarray}
The lowest component gives the 
BPS equation (\ref{BPSeq-Sigma}) for vector multiplet 
scalar $\Sigma$ with $H^2=0$. 
The BPS equation for the second hypermultiplet $H^2$ 
does not follow from the above constraint equation. 
This is natural since our choice of the preserved four supercharges 
implies $H^2\equiv 0$ as the solution of BPS equations. 
By comparing the coefficients of 
$\theta\sigma^\mu\bar\theta$ in 
Eqs.(\ref{eq:constr-vector-comp2}) and 
(\ref{eq:constr-vector-comp1}), 
we find that the vector component of 
Eq.(\ref{eq:superfield-constr1}) precisely gives the 
field equation (\ref{eq:EOM-gauge-field}) 
for the gauge field $W_\mu$ 
which has to be imposed to obtain 
the configuration of the gauge field $W_{\mu}$. 
Thus we see that all the necessary informations to determine 
the field configurations including fluctuations 
are contained systematically in this superfield formulation. 
It is worth empasizing that even 
the BPS equations follow without using 
the usual method like the Bogomol'nyi completion, 
and that it is just an automatic consequence of the 
$\lambda$ expansion with four manifest SUSY.

So far we have not specified any gauge of the 
complexified $U(N_{\rm C})$ local gauge invariance 
in Eqs.(\ref{eq:gauge-tr-vector}) 
and (\ref{eq:gauge-tr-hyper}). 
In the spirit of our method in solving BPS equations, 
we can first solve the constraint equation 
(\ref{eq:superfield-constr2}) for the 
hypermultiplet, and reformulate the rest into a gauge 
invariant fashion. 
Let us define an element of the complexified gauge 
transformation ${\bf S}$ to express the chiral scalar 
superfield ${\bf \Phi}$ for the adjoint scalar of the 
vector multiplet as a pure gauge 
\begin{eqnarray}
{\bf \Phi }={\bf S}^{-1}\partial _y{\bf S}. 
\label{eq:phi-pure-gauge}
\end{eqnarray}
Then the constraint equation 
(\ref{eq:superfield-constr2}) 
for the hypermultiplet chiral 
superfield becomes simpler 
\begin{eqnarray}
 \partial _y({\bf SH}^1)={\bf SH}^1 M ,
\end{eqnarray}
which is easily solved in terms of the moduli 
matrix chiral superfields ${\bf H}_0$ as 
\begin{eqnarray}
{\bf H}^1(x,\theta, \bar \theta,y)
={\bf S}^{-1}(x,\theta, \bar \theta,y)
{\bf H}_0(x,\theta, \bar \theta)e^{My}. 
\label{eq:moduli-matrix-superfield}
\end{eqnarray}
This solution clearly shows that the chiral superfield ${\bf S}$ 
transforms under the complexified $U(N_{\rm C})$ 
gauge transformations in Eqs.(\ref{eq:gauge-tr-vector}) 
and (\ref{eq:gauge-tr-hyper}) as 
\begin{eqnarray}
{\bf S} \rightarrow {\bf S}'= {\bf S} e^{\bf \Lambda}. 
\label{eq:s-gauge-trans}
\end{eqnarray}
On the other hand, there is an ambiguity to separate 
 the chiral superfield ${\bf S}^{-1}$ from the moduli 
matrix chiral superfield 
${\bf H}_0$ in Eq.(\ref{eq:moduli-matrix-superfield}). 
If two sets of chiral superfields $({\bf S}, {\bf H}_0)$ and 
 $({\bf S}', {\bf H}_0')$ are related by 
 an element of ${\cal V}$ of $GL (N_{\rm C},{\bf C})$
\begin{eqnarray}
{\bf S}' = {\cal V}{\bf S},\quad 
{\bf H}_0{}'={\cal V}{\bf H}_0 ,
\label{art-sym}
\end{eqnarray} 
they give identical result for physical quantities such 
as hypermultiplet chiral superfield ${\bf H}^1$. 
Thus the $V$-equivalence transformations \cite{INOS1}, 
\cite{INOS2} in Eq.(\ref{eq:v_equivalence}) are 
supersymmetrized by the chiral superfield ${\cal V}$.

After solving the hypermultiplet constraint equation 
(\ref{eq:superfield-constr2}), 
we can now define a vector 
superfield ${\bf \Omega}$ which is 
invariant under the complexified $U(N_{\rm C})$ gauge 
transformations 
\begin{eqnarray}
{\bf \Omega }
\equiv {\bf S}e^{-2{\bf V}}{\bf S}^\dagger .
\label{eq:def-Omega}
\end{eqnarray}
The remaining constraint equation 
(\ref{eq:superfield-constr1}) 
can be rewritten in terms of the gauge invariant 
superfield ${\bf \Omega}$ as 
\begin{eqnarray}
\partial _y\left({\bf \Omega }^{-1}
\partial _y{\bf \Omega }\right)
=g^2c\left(1-{\bf \Omega }^{-1}{\bf \Omega }_0\right),
\quad
 {\bf \Omega }_0\equiv 
c^{-1}{\bf H}_0e^{2My}{\bf H}_0^\dagger , 
\label{eq:master-eq-superfield} 
\end{eqnarray}
which gives the master equation (\ref{eq:master-eq}) 
as the lowest component. 
Therefore this is the superfield extension of 
the master equation. 

By using the solution of the 
constraint equation (\ref{eq:superfield-constr2}) 
for the hypermultiplet superfield ${\bf H}^1$, 
we can rewrite the fundamental Lagrangian 
in Eq.(\ref{eq:Lag-order2}) 
(up to order ${\cal O}(\lambda^2)$) 
in terms of the gauge invariant 
superfield ${\bf \Omega}$ as 
\begin{eqnarray}
{\cal L}&\!\!\!=&\!\!
-{\cal E}_{\rm w}
+\int d^4 \theta 
\left[
c\log{\rm det}{\bf \Omega} 
+c{\rm Tr}\left({\bf \Omega} _0{\bf \Omega} ^{-1}\right)
+{1\over 2g^2}
{\rm Tr}\left({\bf \Omega }^{-1}
\partial _y{\bf \Omega} \right)^2 
\right]
+{\cal O}(\lambda ^4). 
\label{eq:Lag-omega}
\end{eqnarray}
The first, second, and third terms 
in the $d^4 \theta$ integrand 
come from the corresponding terms in the 
$d^4 \theta$ integrand of the fundamental Lagrangian 
(\ref{eq:Lag-order2}) (up to order ${\cal O}(\lambda^2)$). 
If we apply variational principle to find the minimum of 
the above five-dimensional Lagrangian 
(\ref{eq:Lag-omega}) written in terms of ${\bf \Omega}$, 
we obtain the master equation (\ref{eq:master-eq-superfield}) 
in terms of superfield. 
Since the fundamental Lagrangian (\ref{eq:Lag-order2}) 
is the result of solving the hypermultiplet 
constraint (\ref{eq:superfield-constr2}), 
it is natural to 
expect that the master equation 
(\ref{eq:master-eq-superfield}) is obtained 
by the action principle applied to 
the Lagrangian (\ref{eq:Lag-order2}) 
in terms of ${\bf \Omega}$ 
as the dynamical variable.

The superspace extension (\ref{eq:master-eq-superfield}) 
of the master equation  provides a method to determine all 
the quantities of interest as a systematic expansion in powers 
of Grassmann coordinates $\theta, \bar \theta$ as follows. 
Suppose we have an exact solution 
$\Omega _{\rm sol}(H_0,H_0^\dagger ,y)$ 
for the master equation (\ref{eq:master-eq}) 
as a function of moduli matrix $H_0, H_0^\dagger$ 
 \begin{eqnarray}
  \Omega ={\bf \Omega }\Big|_{\theta =0}
=\Omega _{\rm sol}(H_0(x),H_0^\dagger (x),y). 
 \end{eqnarray}
By promoting the moduli matrix to a superfield 
${\bf H}_0, {\bf H}^\dagger_0$ 
defined in Eq.(\ref{eq:moduli-matrix-superfield}), 
we obtain the solution for the vector superfield ${\bf \Omega} $ 
of the superfield master equation 
(\ref{eq:master-eq-superfield}) as a composite of the chiral and the
anti-chiral superfields, 
\begin{eqnarray}
\Omega _{\rm sol}({\bf H}_0(x,\theta), 
{\bf H}_0^\dagger (x,\bar{\theta}),y)
\equiv 
{\bf \Omega }_{\rm sol}
. 
\end{eqnarray}
As we noted in Eq.(\ref{eq:def-Omega}), 
the superfield ${\bf \Omega}=
{\bf S}e^{-2{\bf V}}{\bf S}^\dagger$ is $U(N_{\rm C})$ 
supergauge invariant, but the division between ${\bf S}$, 
(${\bf S}^\dagger$) and $e^{-2{\bf V}}$ depends on the 
gauge choice. 
In obtaining the solution for the fluctuation fields such 
as $W_\mu$, we need to choose to use the Wess-Zumino 
gauge for the real general (vector) superfield 
${\bf V}_{\rm sol}$. 
This gauge transformation to the Wess-Zumino gauge 
is expressed as a multiplication of 
the chiral ${\bf S}_{\rm sol}$ and anti-chiral 
${\bf S}_{\rm sol}^\dagger$ superfields 
from left and right respectively as 
\begin{eqnarray}
{\bf S}_{\rm sol}e^{-2{\bf V}_{\rm sol}}
{\bf S}^\dagger_{\rm sol}
= 
{\bf \Omega }_{\rm sol} . 
\label{eq:sol-superfield}
\end{eqnarray} 
Then expansion of the left-hand side of 
Eq.(\ref{eq:sol-superfield}) 
in powers of the Grassmann coordinates 
$\theta, \bar \theta$ gives 
\begin{eqnarray}
{\bf S}_{\rm sol}e^{-2{\bf V}_{\rm sol}}
{\bf S}^\dagger_{\rm sol}
= 
S_{\rm sol}S^\dagger _{\rm sol}
+
\theta\sigma^\mu\bar \theta \left(
i(\partial _\mu S_{\rm sol})S_{\rm sol}^\dagger 
-iS_{\rm sol}(\partial _\mu S_{\rm sol}^\dagger )
+2S_{\rm sol}W^{\rm sol}_\mu S_{\rm sol}^\dagger 
\right) +\cdots,
\label{eq:omega-vector}
\end{eqnarray} 
where we have not displayed the bilinear terms of 
fermions, and dots denote other powers of Grassmann 
coordinates. 
Expanding the right-hand side of 
Eq.(\ref{eq:sol-superfield}) we obtain 
\begin{eqnarray}
\Omega _{\rm sol}({\bf H}_0,{\bf H}_0^\dagger ,y)
&\!\!\!=&\!\!\!
\Omega_{\rm sol}
+\theta\sigma^\mu\bar \theta \left(
i(\delta _\mu -\delta _\mu ^\dagger )
\Omega _{\rm sol}\right)
+\cdots , 
\label{eq:omega-sol-vector}
\end{eqnarray}
we define the variation $\delta_\mu$ and its conjugate 
$\delta_\mu^\dagger$ with respect to 
the scalar fields of chiral superfields and 
anti-chiral superfields 
\begin{eqnarray}
\delta_\mu 
\equiv \sum_i\partial_\mu\phi^i
{\delta 
\over\delta \phi^i}, 
\qquad 
\delta_\mu^\dagger 
\equiv \sum_i\partial_\mu \phi^{i*}
{\delta 
\over \delta \phi^{i*}}, 
\label{eq:variation}
\end{eqnarray}
respectively. 
If the variation $\delta_\mu$ and $\delta^\dagger_\mu$ act 
on those functions which depend on the world-volume 
coordinates $x^\mu$ only through moduli fields, they 
satisfy 
\begin{eqnarray}
\partial _\mu =\delta _\mu +\delta _\mu ^\dagger .
\label{eq:variation-sum}
\end{eqnarray}

Comparing the lowest component of (\ref{eq:omega-vector}) 
and (\ref{eq:omega-sol-vector}), we obtain 
\begin{eqnarray}
S_{\rm sol}S^\dagger _{\rm sol}
=
\Omega _{\rm sol}
.
\label{eq:sol-omega}
\end{eqnarray} 
This shows that we cannot avoid 
$S_{\rm sol}$ 
to depend on both $\phi^i$ and $\phi^{i*}$, 
since we cannot factorize these dependences 
in $\Omega_{\rm sol}$. 
One should note that ${\bf S}_{\rm sol}$ 
(${\bf S}_{\rm sol}^\dagger$) is still chiral (anti-chiral) 
scalar superfield, taking both $\phi^i$ and $\phi^{i*}$ 
as lowest components of chiral scalar superfields. 
Comparison of the vector component of (\ref{eq:omega-vector}) 
and (\ref{eq:omega-sol-vector}), we obtain 
a solution of the gauge fields as 
\begin{eqnarray}
- iW_\mu ^{\rm sol}=S^{-1}_{\rm sol}\delta_\mu^\dagger S_{\rm sol}
+S^\dagger _{\rm sol}\delta _\mu S^{\dagger -1}_{\rm sol}
+(\hbox{bi-linear terms of fermions}).
\label{eq:sol-W}
\end{eqnarray}
It is interesting to observe that this solution of gauge 
field fluctuation $W_\mu ^{\rm sol}$ receives contributions 
only from the $\phi^{i*}$ ($\phi^i$) 
dependence of $S_{\rm sol}$ ($S_{\rm sol}^\dagger$), 
in spite of the ${\bf S}_{\rm sol}$ being the chiral 
superfield. 
Similarly the adjoint scalar $\Sigma$ and the gauge 
field $W_y$ in the extra fifth direction is obtained from 
the lowest component of Eq.(\ref{eq:phi-pure-gauge}) 
\begin{eqnarray}
{\bf \Phi }_{\rm sol}
={\bf S}_{\rm sol}^{-1}\partial _y{\bf S}_{\rm sol}
\quad \rightarrow \quad 
\Sigma^{\rm sol} +iW_y^{\rm sol} 
=S^{-1}_{\rm sol}\partial _yS_{\rm sol} .
\end{eqnarray}
The other components of the superfields ${\bf \Omega }$, 
${\bf V}$, and ${\bf \Phi }$ 
are similarly determined by the superfield equations.

In order to obtain the low-energy effective Lagrangian 
${\cal L}_{\rm eff}$ finally, we just need to substitute 
the solutions ${\bf \Omega}_{\rm sol}$ into the fundamental 
Lagrangian ${\cal L}$ and integrate over the extra 
dimensional coordinate $y$. 
The resulting four-dimensional effective Lagrangian for the 
moduli matrix superfield ${\bf H}_0$ is given by  
\begin{eqnarray}
{\cal L}_{\rm eff}&\!\!\!=&\!\!\!
\int dy \; {\cal L}
=-T_{\rm w}
+\int d^4\theta K({\bf \phi} ,{\bf \phi} ^*)+{\cal O}(\lambda ^4),
\label{eq:effective-Lag}
\end{eqnarray}
where the K\"ahler potential is expressed by an integral 
form as 
\begin{eqnarray}
 K({\bf \phi} ,{\bf \phi} ^*)&\!\!\!=&\!\!\!\int dy 
{\cal K}({\bf \phi} ,{\bf \phi} ^*,{\bf \Omega},y)
\Big|_{{\bf \Omega} ={\bf \Omega} _{\rm sol} },
\label{eq:kahler-pot}
\end{eqnarray}
with a density
\begin{eqnarray}
{\cal K}({\bf \phi} ,{\bf \phi} ^*,{\bf \Omega},y)
&\!\!\!=&
c\log{\rm det}{\bf \Omega} 
+c{\rm Tr}\left({\bf \Omega} _0{\bf \Omega} ^{-1}\right)
+{1\over 2g^2}
{\rm Tr}\left({\bf \Omega }^{-1}
\partial _y{\bf \Omega} \right)^2. 
\label{eq:kahler-density}
\end{eqnarray}
Since we are considering the massless fields corresponding to 
the moduli, we naturally obtain a nonlinear sigma model 
whose kinetic terms are specified by the K\"ahler potential 
$K(\phi, \phi^*)$, without any potential terms. 
Let us note that our method gives the density of the K\"ahler 
potential directly without going through the K\"ahler metric. 
This is in contrast to the component approach where one usually 
obtains the K\"ahler metric of the nonlinear sigma model with 
component scalar fields, such as in 
Eq.(\ref{eq:kahler-metric-wall}), 
and then integrate it to obtain 
the K\"ahler potential with a lot of labor.

When we consider a composite state of domain walls, 
vortices (or lumps) and monopoles \cite{INOS3}, 
vortices and monopoles can be interpreted 
in the effective Lagrangian on the domain wall as follows:
the first contribution corresponds to the energy 
of vortices (or lumps), and the third contribution 
corresponds to the energy 
of monopoles. 
It is interesting to note that 
our effective Lagrangian 
is not just an effective Lagrangian on a single wall, but an 
effective Lagrangian on the multiple wall system with 
various moduli such as relative distance moduli as 
the effective fields. 
Therefore we can discuss lumps stretched between multiple 
walls, which was difficult previously \cite{bion}.

By using the superfield master equation 
(\ref{eq:master-eq-superfield}), 
we can show that 
the second term in Eq.(\ref{eq:kahler-density}) becomes 
a total derivative term. 
Therefore it can be omitted from the 
effective Lagrangian. 
The wall tension $T_{\rm w}$ is given by the topological charge 
as an integral over the total derivative term ${\cal E}_{\rm w}$ 
in Eq.(\ref{eq:total-der}) 
\begin{eqnarray}
T_{\rm w}=
\int dy \; {\cal E}_{\rm w}
={\rm Tr}\big[c\Sigma 
\big]^{y=\infty }_{y=-\infty }, 
\end{eqnarray}
where we have used the boundary condition which requires that 
vacua are reached at both infinities $y=\pm \infty$. 

If we take the strong coupling limit $g^2\rightarrow \infty $, 
we find that the superfield master equation 
(\ref{eq:master-eq-superfield}) becomes just an algebraic 
equation $\Omega =\Omega _0$. 
Therefore exact solutions for 
$\Omega $ can be obtained and the K\"ahler potential 
assumes a simple form in this 
case \cite{INOS1} 
\begin{eqnarray}
  K_0(\phi ,\phi ^*)=c\int dy \log{\rm det}\Omega _0. 
\label{eq:kahler-pot-strong}
\end{eqnarray}

\section{Superfield Effective Lagrangian on Vortices}
\label{sc:vortex}

As another example to illustrate the power of our method, 
we shall consider vortex as another $1/2$ BPS soliton in 
this section. 
If we wish to obtain four-dimensional world-volume on 
vortices, we need to use a fundamental theory 
in six dimensions. 
Since this is the theory with eight SUSY in highest 
dimensions, all the lower-dimensional theories can be 
obtained by a simple dimensional reduction and/or 
a Scherk-Schwarz dimensional reduction with twisted 
boundary conditions along the compactified directions. 
We take again the $U(N_{\rm C})$ supersymmetric gauge theory 
with $N_{\rm F}$ (massless) hypermultiplets in the 
fundamental representation.\footnote{
Hypermultiplets in six-dimensions must be massless. 
} 
Vortex solutions in six dimensions with 
manifest four supercharges in four-dimensional world-volume 
were discussed \cite{Eto:2004ii}. 
The bosonic part of the fundamental Lagrangian in six 
dimensions is given by 
\begin{eqnarray}
{\cal L}|_{\rm boson}
&\!\!\!=&\!\!\! 
{\rm Tr}\biggl[-{1\over 2g^2} F_{MN}(W)F^{MN}(W)
+\frac{1}{g^2}\sum_{a=1}^3(Y^a)^2-cY^3 
\nonumber \\
&\!\!\!-&\!\!\!  {\cal D}^M H^i ({\cal D}_M H^i)^\dagger 
+F^iF^{i\dagger} 
+Y^a(\sigma^a)_{ij}H^j H^{i\dagger} 
\biggr], 
\label{fundamental-Lag2-6d}
\end{eqnarray}
where the indices $M, N=0, 1, \cdots, 5$ run over six 
spacetime dimensions. 
The Lagrangian in five dimensions with massive 
hypermultiplets in Eq.(\ref{fundamental-Lag2}) can be 
obtained by a Scherk-Schwarz dimenisonal reduction 
along the sixth direction $x^5$. 

The fundamental Lagrangian of the 
supersymmetric gauge theories in six dimensions can be written in 
terms of the superfields with four SUSY manifest 
\cite{AGW}. 
We can use the same superfields as in five dimensions in 
Eqs.(\ref{eq:vector-superfield}), 
(\ref{eq:vectorr-chiral-superfield}), 
(\ref{eq:hyper-superfield1}), (\ref{eq:hyper-superfield2}) 
except for two features. 
The component fields should now depend on six dimensions 
rather than five dimensions, and the auxiliary fields for 
eight SUSY $Y^3$ and $F^i$ are slightly more shifted from 
those for four supercharges ${\cal Y}^3$ and ${\cal F}^i$, compared 
to the five-dimensional case in 
Eqs.(\ref{eq:vector-superfield}), 
(\ref{eq:hyper-superfield1}) and 
(\ref{eq:hyper-superfield2}). 
The chiral and anti-chiral superfields ${\bf H}^1$ and 
${\bf H}^2$ for hypermultiplets 
are given by 
\begin{eqnarray}
{\bf H}^1=H^1+\sqrt{2}\theta \psi_++\theta^2 {\cal F}^1, 
\qquad 
{\cal F}^1\equiv F^1+({\cal D}_4+i{\cal D}_5)H^2, 
\label{eq:hyper-superfield1-6}
\end{eqnarray}
\begin{eqnarray}
{\bf H}^2=H^2+\sqrt{2}\bar \theta \bar \psi_-
+\bar \theta^2 {\cal F}^2, 
\qquad 
{\cal F}^2\equiv -F^2-({\cal D}_4-i{\cal D}_5)H^1. 
\label{eq:hyper-superfield2-6}
\end{eqnarray}
The chiral scalar superfield 
${\bf \Phi}$ containing the adjoint scalar is given by 
\begin{equation}
{\bf \Phi }
=W_5 + i W_4 + \sqrt{2}\theta (-i\sqrt{2} \lambda_-) 
+ \theta^2 (Y^1 +i Y^2), 
\label{eq:vectorr-chiral-superfield-6}
\end{equation}
where we identify the adjoint scalar field $\Sigma$ to be 
the sixth component of the gauge field $W_5$ and denote the 
fifth component of gauge field as $W_4$ rather than $W_y$. 
Finally the vector superfield ${\bf V}$ contains 
gauge field, gaugino, and auxiliary field and has 
a decomposition in the Wess-Zumino gauge 
\begin{equation}
{\bf V}\Big|_{\rm WZ}=
-\theta \sigma^\mu \bar\theta W_\mu 
+ i\theta^2\bar \theta \bar \lambda_+
- i\bar \theta^2 \theta \lambda_+ 
+{1 \over 2} \theta^2\bar \theta^2 {\cal Y}^3, 
\qquad {\cal Y}^3\equiv Y^3-F_{45}(W).  
\label{eq:vector-superfield-6}
\end{equation}

Let us define complex variables for extra two dimensions 
\begin{eqnarray}
z\equiv x^4 +ix^5, 
\qquad 
\partial \equiv {1 \over 2}(\partial_4-i\partial_5).  
\end{eqnarray}
The six-dimensional Lagrangian can be rewritten in terms of 
these superfields for four SUSY. 
It contains a term ${\cal L}_{\rm sym}$ similar to 
that in the five-dimensional case 
\begin{eqnarray}
 {{\cal L}_{\rm sym}}
&\!\!\!
=
&\!\!\!
\int d ^4\theta {\rm Tr}\biggl[
-2c{\bf V}
+{1\over g^2}\left(
2e^{-2{\bf V}}{\bar \partial}e^{2{\bf V}}
e^{-2{\bf V}}{\partial}e^{2{\bf V}}
-2{\partial}e^{2{\bf V}}e^{-2{\bf V}}{\bf \Phi}^\dagger 
\right.
\nonumber \\
&\!\!\!&\!\!\!
{}
\left.-2e^{-2{\bf V}}{\bar \partial}e^{2{\bf V}}{\bf \Phi} 
+e^{-2{\bf V}}{\bf \Phi}^\dagger e^{2{\bf V}}{\bf \Phi}
\right)
+e^{2{\bf V}}{\bf H}^1{\bf H}^{1\dagger }
+e^{-2{\bf V}}{\bf H}^2{\bf H}^{2\dagger }
\biggr]
\nonumber \\
&\!\!\!&\!\!\!
{}+\left(\int d ^2\theta 
{\rm Tr}\left[
(2{\partial}+{\bf \Phi}){\bf H}^1{\bf H}^{2\dagger }
+{1\over 4g^2}W^\alpha W_\alpha 
\right]+{\rm h.c.}\right). 
\label{eq:fund-Lag-6d-superfield}
\end{eqnarray}
However, it has been known that the fundamental Lagrangian 
with eight SUSY in six-dimensions requires an addition of a 
Wess-Zumino-Witten like term ${\cal L}_{\rm WZW}$ in order 
to maintain the complexified $U(N_{\rm C})$ gauge 
invariance if superfields for four SUSY are used 
\cite{AGW}, \cite{MSS} 
\begin{eqnarray}
 {\cal L}_{\rm WZW}
\equiv \int d^4\theta {\rm Tr}\left[{16 \over g^2} 
\bar\partial {\bf V} 
{\sinh(2L_{\bf V})-2L_{\bf V}\over (2L_{\bf V})^2}
\partial {\bf V}\right]
, 
\label{eq:WZW-term}
\end{eqnarray}
where the operation $L_{\bf V}$ is defined by 
\begin{eqnarray}
L_{\bf V}\times {\bf X}=[{\bf V},\,{\bf X}]. 
\label{eq:adjoint-operation}
\end{eqnarray}
To rewrite the fundamental six-dimensional Lagrangian 
${\cal L}$ in Eq.(\ref{fundamental-Lag2-6d}) in 
terms of the superfields for four SUSY, we need to retain 
total divergence terms 
$-{\cal E}_{\rm v}$ 
when we make 
partial integrations in the codimensions of the soliton, 
namely fifth and sixth directions, 
similarly to the wall case in five dimensions 
\begin{eqnarray}
 {\cal L}=-{\cal E}_{\rm v}
+ {\cal L}_{\rm sym} 
+ {\cal L}_{\rm WZW}. 
\end{eqnarray}
It is easiest to evaluate the total divergence term 
${\cal E}_{\rm v}$ in the Wess-Zumino gauge 
(\ref{eq:vector-superfield-6}) where 
the Wess-Zumino-Witten like term in Eq.(\ref{eq:WZW-term}) 
vanishes ${\cal L}_{\rm WZW}=0$. 
We also need to take account of the difference of 
auxiliary fields for four supercharges ${\cal Y}^3$, 
and ${\cal F}^i$ from those for eight SUSY $Y^3$, $F^i$ 
in Eqs.(\ref{eq:vector-superfield-6}), 
(\ref{eq:hyper-superfield1-6}),  and 
(\ref{eq:hyper-superfield2-6}). 
We obtain the total divergence term ${\cal E}_{\rm v}$ 
representing the topological charge contributing to 
the energy density of the (vortex) background which maintains 
four SUSY, exhibiting only the bosonic terms explicitly as 
\begin{eqnarray}
{\cal E}_{\rm v}
&\!\!\!=&\!\!{\rm Tr}\big[c F_{45}(W) 
-{2\over g^2}\left(\partial_4({\cal Y}^3W_5) 
-\partial_5({\cal Y}^3W_4)\right) 
+2\partial({\cal F}^1H^{2\dagger })
+2\bar \partial(H^2{\cal F}^{1\dagger })
\nonumber \\
&\!\!\!&\!\!\! 
+i\left(\partial_4(H^{1\dagger}{\cal D}_5H^1
-H^{2\dagger}{\cal D}_5H^2)
-\partial_5(H^{1\dagger}{\cal D}_4H^1
-H^{2\dagger}{\cal D}_4H^2)\right)
\big]
.
\label{eq:total-der-6}
\end{eqnarray}

Let us observe that only the first term contributes to the 
energy when the total divergence term is integrated over the 
two-dimensional plane, since we require that the field 
configuration should reduce to vacuum at spacial infinity 
where four SUSY are maintained: ${\cal Y}^3=0$, and 
${\cal F}^i=0$.

Although the Wess-Zumino gauge is useful to reveal the 
physical field content, it is more useful to consider 
generic gauges in order to perform a systematic expansion in 
powers of the slow-movement parameter $\lambda$. 
If we do not choose the Wess-Zumino gauge, we need to retain 
the Wess-Zumino-Witten like term. 
As shown in Appendix \ref{sc:variation-6d-Lag}, 
the first term ($2e^{-2{\bf V}}{\bar \partial}e^{2{\bf V}}
e^{-2{\bf V}}{\partial}e^{2{\bf V}}$) 
in the bracket (multiplied by $1/g^2$) 
of Eq.(\ref{eq:fund-Lag-6d-superfield}) can be combined with 
the Wess-Zumino-Witten-like term (\ref{eq:WZW-term}) to give the 
following total fundamental Lagrangian as
\begin{eqnarray}
 {\cal L}
&\!\!\!
=
&\!\!\!
-{\cal E}_{\rm v} + 
\int d ^4\theta {\rm Tr}\biggl[
-2c{\bf V}
+{1\over g^2}\bigl(16\int_0^1 dx\int_0^x dy
{\bar \partial}{\bf V} e^{2yL_{\bf V}}
{\partial}{\bf V}
-2{\partial}e^{2{\bf V}}e^{-2{\bf V}}{\bf \Phi}^\dagger 
\nonumber \\
&\!\!\!&\!\!\!
{}
-2e^{-2{\bf V}}{\bar \partial}e^{2{\bf V}}{\bf \Phi} 
+e^{-2{\bf V}}{\bf \Phi}^\dagger e^{2{\bf V}}{\bf \Phi}
\bigr)
+e^{2{\bf V}}{\bf H}^1{\bf H}^{1\dagger }
+e^{-2{\bf V}}{\bf H}^2{\bf H}^{2\dagger }
\biggr]
\nonumber \\
&\!\!\!&\!\!\!
{}+\left(\int d ^2\theta 
{\rm Tr}\left[
(2{\partial}+{\bf \Phi}){\bf H}^1{\bf H}^{2\dagger }
+{1\over 4g^2}W^\alpha W_\alpha 
\right]+{\rm h.c.}\right). 
\label{eq:fund-Lag-combined}
\end{eqnarray}

Let us find out the order of $\lambda$ for various fields. 
It is well-known that the first hypermultiplet $H^1$ 
has non-vanishing values, whereas the second hypermultiplet 
$H^2$ vanishes \cite{Hanany:2003hp}--\cite{vortices}. 
When we promote the moduli parameters to fields on the 
world volume of the soliton, the fluctuation naturally 
induces terms of order 
$\lambda$  to both $H^1, H^2$. 
Therefore we need to assume 
the same order assignment for various fields as in the wall case 
(\ref{BPS-cond-by-order-assign-wall-1}), 
(\ref{BPS-cond-by-order-assign-wall-2}), and 
(\ref{BPS-cond-by-order-assign-wall-3}). 
Now let us expand the fundamental Lagrangian in powers of 
the slow-movement parameter $\lambda$. 
If we retain terms up to the order of ${\cal O}(\lambda^2)$, 
we obtain 
\begin{eqnarray}
 {\cal L}
&=& 
-{\cal E}_{\rm v} + 
\int d ^4\theta {\rm Tr}\biggl[
-2c{\bf V} 
+{1\over g^2}\bigl(16\int_0^1 dx\int_0^x dy
{\bar \partial}{\bf V} e^{2yL_{\bf V}}
{\partial}{\bf V}
-2{\partial}e^{2{\bf V}}e^{-2{\bf V}}{\bf \Phi}^\dagger 
\nonumber \\
&\!\!\!&\!\!\!
{}
-2e^{-2{\bf V}}{\bar \partial}e^{2{\bf V}}{\bf \Phi} 
+e^{-2{\bf V}}{\bf \Phi}^\dagger e^{2{\bf V}}{\bf \Phi}
\bigr)
+e^{2{\bf V}}{\bf H}^1{\bf H}^{1\dagger }
\biggr]
\nonumber \\
&\!\!\!&\!\!\!
{}
+\left(\int d ^2\theta 
{\rm Tr}\left[
(2{\partial}+{\bf \Phi}){\bf H}^1{\bf H}^{2\dagger }
\right]+{\rm h.c.}\right)
+{\cal O}(\lambda^4). 
\label{eq:vortex-expansion}
\end{eqnarray}
The superfield ${\bf H}^2$ now acts as a Lagrange multiplier. 
The constraint resulting from varying ${\bf H}^2$ gives 
the BPS equation for hypermultiplet ${\bf H}^1$ in 
the superfield form 
\begin{eqnarray}
(2{\partial}+{\bf \Phi}){\bf H}^1=0. 
\label{eq:superfield-constr3}
\end{eqnarray}
This BPS equation can be easily solved in terms of a 
complexified gauge transformation superfield 
${\bf S}$ defined as 
\begin{eqnarray}
{\bf \Phi}=2{\bf S}^{-1}\partial{\bf S}, 
\qquad 
{\bf H}^1(x^\alpha, \theta, \bar \theta, z, \bar z) 
={\bf S}^{-1}(x^\alpha, \theta, \bar \theta, z, \bar z)
{\bf H}_0(x^\alpha, \theta, \bar \theta, \bar z). 
\end{eqnarray}
Similarly to the wall case, we can define a gauge invariant 
vector superfield ${\bf \Omega}$ as 
\begin{eqnarray}
{\bf \Omega}(x^\alpha, \theta, \bar \theta, z, \bar z) 
\equiv {\bf S}(x^\alpha, \theta, \bar \theta, z, \bar z)
e^{-2{\bf V}}
{\bf S}^\dagger (x^\alpha, \theta, \bar \theta, z, \bar z)
. 
\end{eqnarray}
Another Lagrange multiplier superfield ${\bf V}$ in 
Eq.(\ref{eq:vortex-expansion}) gives another constraint 
\begin{eqnarray}
&\!\!\!&\!\!\!
g^2(c-{\bf H}^1{\bf H}^{1\dagger }e^{2{\bf V}}) 
=
- 2\partial\left(e^{-2{\bf V}}2\bar\partial e^{2{\bf V}}
- 2{\bf \Phi}^\dagger e^{2{\bf V}} \right)
- 2\bar\partial{\bf \Phi} 
\nonumber \\
&\!\!\!&\!\!\!+ {\bf \Phi}\left(e^{-2{\bf V}}2\bar\partial e^{2{\bf V}}
- 2{\bf \Phi}^\dagger e^{2{\bf V}} \right)
-\left(e^{-2{\bf V}}2\bar\partial e^{2{\bf V}}
- 2{\bf \Phi}^\dagger e^{2{\bf V}} \right) {\bf \Phi} .
\label{eq:superfield-constr4}
\end{eqnarray} 
This superfield constraint contains the BPS equation for 
vector multiplet as the lowest component (independent of 
the Grassmann coordinate $\theta$) 
\begin{eqnarray}
g^2(c- H^1 H^{1\dagger }) 
&\!\!\!
=
&\!\!\!
 2F_{45}(W) .
\label{eq:vortex-BPSeq-vector}
\end{eqnarray} 

We can rewrite the constraint equation in terms of the 
gauge invariant superfield $\Omega$ as 
\begin{eqnarray}
g^2c(1-{\bf \Omega}_0{\bf \Omega}^{-1}) 
&\!\!\!
=
&\!\!\!
 2\partial\left(2\bar\partial {\bf \Omega} {\bf \Omega}^{-1}
 \right), 
\label{eq:master-eq-vortex}
\end{eqnarray}
whose lowest component (independent of the Grassmann 
coordinate $\theta$) is precisely the master equation 
in the case of vortices \cite{Eto:2005yh}.\footnote{ 
The master equation reduces to the so-called 
Taubes equation \cite{Taubes:1979tm} 
in the case of Abrikosov-Nielsen-Olesen vortices 
\cite{Abrikosov:1956sx,Nielsen:cs} ($N_{\rm C} = N_{\rm F} = 1$)
by rewriting 
$c \Omega(z,\bar z)=|H_0|^2 e^{-\xi (z,\bar z)}$ 
with $H_0 = \prod_i(z-z_i)$. 
Note that $\log \Omega$ 
is regular everywhere 
while $\xi$ is singular at vortex positions.
}
Changing variables from ${\bf \Phi}$ to ${\bf S}$ is 
nothing but a complexified $U(N_{\rm C})$ gauge 
transformation to choose a gauge where ${\bf \Phi}$ is 
eliminated. 
In that gauge, we obtain 
$e^{-2{\bf V}}\rightarrow {\bf \Omega}$, 
${\bf H}^1\rightarrow {\bf H}_0$. 

In order to obtain the effective Lagrangian up to two 
derivatives, we solve the constraint equation 
(\ref{eq:superfield-constr3}) for hypermultiplet 
 superfield ${\bf H}^1$ and use the solution to rewrite 
the fundamental Lagrangian in 
Eq.(\ref{eq:vortex-expansion}) 
(up to order ${\cal O}(\lambda^2)$) 
in terms of the gauge invariant 
superfield ${\bf \Omega}$. 
After integrating over  two extra dimensions $x^4, x^5$, 
we obtain the effective Lagrangian for moduli fields 
whose nonlinear kinetic term is described in 
terms of the K\"ahler potential $K$ apart from the energy 
of the background $1/2$ BPS vortices $T_{\rm v}$ 
\begin{eqnarray}
{\cal L}_{\rm eff}&\!\!\!=&\!\!
\int dx^4 dx^5 \; {\cal L}
=-T_{\rm v}
+\int d\theta ^4K({\bf \phi} ,{\bf \phi} ^*)+{\cal O}(\lambda ^4), 
\label{eq:effective-Lag-vortex}
\end{eqnarray}
\begin{eqnarray}
 T_{\rm v}
= \int d x^4 dx^5 {\cal E}_{\rm v}
= \int d x^4 dx^5 {\rm Tr}\big[c F_{45}(W)\big]. 
\label{eq:vortex-energy}
\end{eqnarray}
To obtain the K\"ahler potential we should evaluate the 
following general formula 
\begin{eqnarray}
 K&=&\int d x^4 d x^5
{\rm Tr}\left[-2c{\bf V} 
+e^{2 {\bf V}}{\bf H}_0{\bf H}_0^\dagger 
+{16\over g^2}\int ^1_0dx\int ^x_0dy
\bar\partial {\bf V}e^{2yL_{\bf V}}\partial {\bf V}\right], 
\end{eqnarray}
where the integration over the  $x^4$-$x^5$ plane may require 
regularization. 
However, we believe that the divergent pieces can only come 
from two possible sources: 
the first possibility is the terms that can be gauged 
away as K\"ahler transformations which does not affect the 
physical K\"ahler metric, and the 
second possibility is the non-normalizable modes which 
have support extending to infinity and should be excluded from the 
dynamical variable in the effective Lagrangian after all. 
The K\"ahler metric \cite{Samols:1991ne} and 
its potential \cite{Chen:2004xu} for the Abelian gauge 
theroy has been obtained before. 
Let us again emphasize that the above formula for the K\"ahler 
pontential density is obtained without the need of computing 
the K\"ahler metric first in contrast to the component approach. 

In order to obtain the K\"ahler metric explicitly, 
we just need to vary this K\"ahler potential. 
In varying the K\"ahler potential, one should be careful 
to interchange the variation and the integration over 
the $x^4$-$x^5$ plane\footnote{
We thank Nick Manton for explaining this point to us. 
}, because of the possible necessity 
of regularization \cite{Chen:2004xu}. 
Assuming the variation can be interchanged with the integration, 
the variation of this K\"ahler potential of the effective 
Lagrangian can be expressed in terms of the gauge invariant 
superfield ${\bf \Omega}=e^{-2{\bf V}}$ as 
\begin{eqnarray}
\delta K&=&\int d x^4 dx^5{\rm Tr}\left[\Omega ^{-1}
\delta \Omega 
\left\{c({\bf 1}_{N_{\rm C}}-\Omega _0\Omega ^{-1})
-{4\over g^2}\partial \left(\bar 
\partial \Omega \Omega ^{-1}\right)\right\}
+c\Omega ^{-1}\delta \Omega _0\right]. 
\end{eqnarray}
We need to express all the superfields in terms of 
the moduli fields by substituting them with the solution of the 
constraint equation (\ref{eq:master-eq-vortex}), namely the 
master equation. 
Then 
the terms in the curly bracket vanish and the ${\bf \Omega}_0$ 
in the last term can also be rewritten in terms of the solution 
${\bf \Omega}$ as 
\begin{eqnarray}
&\!\!\!&\!\!\!\delta K\Big|_{\Omega =\Omega _{\rm sol}}
=\int dx^4 dx^5
{\rm Tr}\left[
c\Omega ^{-1}\delta \Omega _0\right]\Big|_{\Omega =\Omega _{\rm sol}} 
\nonumber \\
&\!\!\!=&\!\!\!\int dx^4 dx^5
{\rm Tr}\left[
-{4\over g^2}\delta \left\{\partial \left(\bar 
\partial \Omega \Omega ^{-1}\right)\right\}
+\delta \Omega \Omega^{-1}\left\{c 
-{4\over g^2}\partial \left(\bar 
\partial \Omega \Omega ^{-1}\right)\right\}
\right]\Big|_{\Omega =\Omega _{\rm sol}}. 
\end{eqnarray}
By choosing the variation $\delta$ as the 
variation with respect to (the scalar fields in) 
chiral superfields $\delta^\mu$ defined in 
Eq.(\ref{eq:variation}), and by varying once more 
by its conjugate $\delta ^\dagger _\mu$ with 
respect to anti-chiral superfields, 
we can express the resulting K\"ahler metric more explicitly 
in terms of the gauge invariant quantity $\Omega$ appearing 
in the master equation (\ref{eq:master-eq-vortex}) 
\begin{eqnarray}
\delta ^{\dagger \mu }\delta _\mu  K
&\!\!\!\!&\!\!\!\!\Big|_{\Omega =\Omega _{\rm sol}}
=\int dx^4 dx^5{\rm Tr}\left[\delta ^{\dagger \mu }\delta _\mu 
c\log\Omega \right.\nn\\
&\!\!\!\!&\!\!\!\!+\left.{4\over g^2}\left\{
\partial \left(\delta ^\mu \Omega \Omega ^{-1}\right)
\delta ^\dagger _\mu 
\left(\bar \partial \Omega \Omega ^{-1}\right)
-\partial (\bar\partial \Omega \Omega ^{-1})
\delta ^\dagger _\mu \left(\delta ^\mu \Omega 
\Omega ^{-1}\right)\right\}\right]
\Big|_{\Omega =\Omega _{\rm sol}}, 
\end{eqnarray}
where we used an identity 
\begin{eqnarray}
 &&\delta ^\dagger \left[\Omega 
\delta \Omega ^{-1}\partial 
\left(\bar \partial \Omega \Omega ^{-1}\right)\right]
+\partial \left[\delta \Omega \Omega ^{-1}
\delta ^\dagger \left(\bar \partial \Omega 
\Omega ^{-1}\right)\right]
\nn\\
&=&\partial \left(\delta \Omega \Omega ^{-1}\right)
\delta ^\dagger \left(\bar \partial \Omega \Omega ^{-1}\right)
-\partial (\bar\partial \Omega \Omega ^{-1})
\delta ^\dagger \left(\delta \Omega \Omega ^{-1}\right), 
\end{eqnarray}
and omitted a surface term 
$\partial \left[\delta \Omega \Omega ^{-1}
\delta ^\dagger \left(\bar \partial \Omega 
\Omega ^{-1}\right)\right]$ by using 
$\delta \Omega \Omega ^{-1}=0$ 
at $x^4, x^5\rightarrow \infty $. 
Except for the conventions differences, this K\"ahler 
metric for the vortex effective theory 
agrees with our previous result in Ref.\cite{Eto:2004rz}.

\section{Effective Lagrangian of Multi-wall System 
\label{sc:eff-Lag-multi-wall}}

In this section, we will perform the integral of the 
formulas (\ref{eq:kahler-pot}) and 
(\ref{eq:kahler-density}) of K\"ahler potential for some 
examples of multi-wall systems to obtain explicit 
effective Lagrangians. 
In this stage, we should remove fake divergences contained 
in the formula by use of the K\"ahler transformation. 
Physical divergences implying non-normalizable modes 
should emerge if there exist moduli 
in vacua of the original Yang-Mills-Higgs system,
whereas we do not consider such cases here.

\subsection{Coordinates for the Moduli Space}
We apply our method to the Abelian gauge theories 
with $N_{\rm F}$ flavors. 
In order to be facilitate an explicit evaluation of the 
integral, let us consider the case where the mass 
differences of hypermultiplets are quantized as 
\begin{eqnarray}
 M={\rm diag}(m_1,\cdots,m_{N_{\rm F}})\simeq 
m\, {\rm diag}(n_1,n_2,\cdots,n_{N_{\rm F}-1},0), 
\quad n_A\in {{\bf Z}_+}, 
\end{eqnarray}
with $n_{A}>n_{A+1},(n_{N_{\rm F}}=0)$. 
By using the $V$-equivalence transformations in 
Eq.(\ref{eq:v_equivalence}) \cite{INOS1}, 
we can choose a parametrization of the moduli matrix 
$H_0$ in terms of $N_{\rm F}-1$ complex moduli parameters 
 $\tau ^A$ as 
\begin{eqnarray}
H_0=\sqrt{c}\left(1,\tau ^2,\tau ^3,\cdots, 
\tau ^{N_{\rm F}}\right),  \quad 
(\tau ^1=1)
\end{eqnarray}
where $\tau ^A\in {\bf C}$ for $2\leq A \leq N_{\rm F}-1$ 
and $\tau ^{N_{\rm F}}\in  {\bf C}-\{0\}$.
The complex moduli parameters are the 
coordinates for the moduli space of 
multi-wall configrations. 
With this moduli matrix parametrization, 
the source term $\Omega _0$ of the master equation 
(\ref{eq:master-eq}) is given by 
\begin{eqnarray}
\Omega _0(y)=
\sum_{A=1}^{N_{\rm F}}|\tau ^{A}|^2w^{n_A}\equiv P(w), 
\quad w\equiv e^{2m y}.
\label{eq:omega0_P} 
\end{eqnarray} 
with a polynomial $P(w)$ of order $n_1$.
Vacua can be characterized by the flavors of the 
nonvanishing hypermultiplets. 
The $N_{\rm F}$ terms $|\tau ^{A}|^2w^{n_A}$ 
in $\Omega_0$ represent weights of $N_{\rm F}$  
vacua, which change as $y$ varies. 
The wall is located at the position where weights of two 
adjacent vacua become equal. 
Thus, the position $y_A$ 
of the wall interpolating the $A$-th vacuum and the 
$(A+1)$-th vacuum is estimated as
\begin{eqnarray}
 y_A\approx -{1\over m}
{\rm Re}{\log\tau ^{A}-\log\tau ^{A+1}\over n_A-n_{A+1}}
\label{eq:positions}
\end{eqnarray}
The moduli parameters $\{\tau ^A\}$  have 
a physical meaning of 
the positions of the wall and the phase of the vacua.

\subsection{Integral at the Strong Gauge Coupling Limit }
By taking the strong gauge coupling limit 
$g^2\rightarrow \infty$, the master equation 
(\ref{eq:master-eq}) can be solved 
algebraically to give the solution 
\begin{eqnarray}
\Omega (y)=\Omega _0(y)=P(e^{2my}). \label{eq:strong_sol}
\end{eqnarray}
As a function of a complex variable 
$w$, the polynomial $P(w)$ has $n_1$ possible complex 
zeros $P(w_k)=0$, and thus 
\begin{eqnarray}
 P(w)=\prod_{k=1}^{n_1}(w-w_k). \label{eq:nzeros}
\end{eqnarray}
By substituting this solution to the formula
(\ref{eq:kahler-pot-strong}) and taking an infrared 
cutoff $2mL$, the K\"ahler potential is rewritten as 
\begin{eqnarray}
K_0=\lim_{L\rightarrow \infty }{c\over 2m}\,
\sum_{k=1}^{n_1}\int ^L_{-L}dt\log(e^{t}-w_k).  
\end{eqnarray}
The integral in the right hand side can be performed 
with the help 
of the following identity 
\begin{eqnarray}
 \int ^L_{-L}dt\log(e^t+e^s)
={1\over 2}(L+s)^2+{\pi \over 6}+{\cal O}(e^{-L}),
\end{eqnarray}
to result in 
\begin{eqnarray}
K_0= \lim_{L\rightarrow \infty }{c\over 2m}\left\{
\sum_{k=1}^{n_1}{1\over 2}[\log(-w_k)]^2+
L\sum_{k=1}^{n_1}\log(-w_k)+n_1
\left({L^2\over 2}+{\pi \over 6}\right)
+{\cal O}(e^{-L})\right\}.\label{eq:prekahler_strong}
\end{eqnarray}
Comparing Eqs.(\ref{eq:omega0_P}) and 
(\ref{eq:nzeros}) we
obtain $\prod_{k=1}^{n_1}(-w_k)=|\tau ^{N_{\rm F}}|^2$, 
that is,
\begin{eqnarray}
 \sum_{k=1}^{n_1}\log(-w_k)
=\log\tau ^{N_{\rm F}}+\log(\tau ^{N_{\rm F}})^*. 
\end{eqnarray}
Therefore, the second and the third terms in right hand 
side of Eq.(\ref{eq:prekahler_strong}) can be 
eliminated by the K\"ahler transformation, and we 
obtain a convergent simple result 
\begin{eqnarray} 
K_0(\tau ^A,\tau ^{A\dagger })
= {c\over 4m}\sum_{k=1}^{n_1}[\log(-w_k)]^2.
\label{eq:kahler_strong}
\end{eqnarray}
Here note that the quantities $w_k$ defined by 
(\ref{eq:nzeros}) are highly non-tribial functions 
of $|\tau ^A|^2$.  
For instance, let us take the case of $N_{\rm F}=3$ with 
$(n_1,n_2,n_3)=(2,1,0)$, and change the parametrization 
of the moduli matrix $H_0$ from $\tau^2, \tau^3$ to 
two complex moduli parameters $\phi _+,\,\phi _-$ as 
\begin{eqnarray}
H_0=\sqrt{c}\left(1,\tau ^2,\tau ^3\right)
=\sqrt{c}\left(1,e^{{\phi _++\phi_-\over 2}},
e^{\phi _+}\right). 
\label{eq:Nf3parametrization}
\end{eqnarray}
We find that the formula (\ref{eq:kahler_strong}) leads 
straightforwardly to 
\begin{eqnarray}
K_0(\phi ,\phi ^*)
={c\over 2m}\left\{\left({\rm Re}(\phi _+)\right)^2
+\left[\log\left({|e^{\phi _-}|
+\sqrt{|e^{\phi _-}|^2-4}\over 2}\right)\right]^2\right\}, 
\end{eqnarray}
where ${\rm Re}\,\phi _+/2m$ and ${\rm Re}\,\phi _-/m$ 
are the center of mass and the relative distance between 
two walls, respectively, according to 
Eq.(\ref{eq:positions}). 
This K\"ahler potential gives precisely the K\"ahler 
metric found by D.Tong\cite{To}. 
Let us note that finding the K\"ahler potential from the 
K\"ahler metric may appear straight-forward, but is 
often nontrivial in reality. 
It is one of the merits of our formulation to obtain the 
K\"ahler potnetial directly without going through 
computation of K\"ahler metric and its integration.

\subsection{
Exact Result at a Finite Gauge Coupling $g^2c= m^2$
}
For particular discrete finite values of gauge coupling, 
we have previously found exact solutions \cite{IOS1} 
of the master equation (\ref{eq:master-eq}). 
Especially, if we choose 
\begin{eqnarray}
g^2c= m^2, 
\label{eq:finite_gauge_coup}
\end{eqnarray}
and $M=m\,{\rm diag}(2,1,0)$, 
we obtain a double-wall solution for full moduli including 
the distance between the walls. 
With the parametrization (\ref{eq:Nf3parametrization}), 
the solution $\Omega$ is given by 
\begin{eqnarray}
 \Omega 
&=&|e^{\frac{\phi _+}{2}}|^2e^{2my}
\left(e^{my}|e^{-{\phi _+\over 2}}|
+e^{-my}|e^{\phi _+\over 2}|
+\sqrt{6+|e^{\phi _-}|}\right)^2. 
\label{eq:omega-finite-g}
\end{eqnarray}

We can evaluate the integral formula 
(\ref{eq:kahler-pot}) and (\ref{eq:kahler-density}) 
for the K\"ahler potential $K(\phi, \phi^*)$ 
by inserting the solution $\Omega$ in 
Eq.(\ref{eq:omega-finite-g}) and by integrating 
over $y$. 
Using the same method as that in the previous 
section, we obtain the first term in 
Eq.(\ref{eq:kahler-density}) explicitly as 
\begin{eqnarray}
c\int ^\infty _{-\infty }dy\log\Omega 
&=&{2c\over m}\left\{{1\over 4}
\left({\rm Re}(\phi _+)\right)^2
+\left[\log\left({\sqrt{6+|e^{\phi _-}|}
+\sqrt{2+|e^{\phi _-}|}\over 2}\right)\right]^2\right\}. 
\end{eqnarray}
The third term in Eq.(\ref{eq:kahler-density}) is found to 
be 
\begin{eqnarray}
{1\over 2g^2}\int ^\infty _{-\infty }dy
(\Omega ^{-1}\partial _y\Omega )^2
&=&
-{4m\over g^2}\sqrt{6+|e^{\phi _-}|\over 2+|e^{\phi _-}|}
\log\left({\sqrt{6+|e^{\phi _-}|}
+\sqrt{2+|e^{\phi _-}|}\over 2}\right), 
\end{eqnarray}
where we used an integration formula 
\begin{eqnarray}
 \int ^L_{-L}dz{e^{2z}\over (e^z+x_1)(e^z+x_2)}=L
+{x_1\log x_1-x_2\log x_2\over x_1-x_2}+{\cal O}(e^{-L}), 
\end{eqnarray}
whereas the contribution from the second term in 
Eq.(\ref{eq:kahler-density}) vanishes. 
Since the gauge coupling $g^2$ is related to the mass 
parameter $m$ by Eq.(\ref{eq:finite_gauge_coup}), 
the K\"ahler potential of the exact solution 
at this finite value of the gauge coupling is 
given by 
\begin{eqnarray}
 K(\phi ,\phi ^*)
&=&{2c\over m}\left\{{1\over 4}
\left({\rm Re}(\phi _+)\right)^2
+\left(\log\left({\sqrt{6+|e^{\phi _-}|}
+\sqrt{2+|e^{\phi _-}|}\over 2}\right)\right)^2
\right.\nn\\
&&\qquad \left.{}
-2\sqrt{6+|e^{\phi _-}|\over 2+|e^{\phi _-}|}
\log\left({\sqrt{6+|e^{\phi _-}|}
+\sqrt{2+|e^{\phi _-}|}\over 2}\right)\right\}. 
\label{eq:kahler_finite} 
\end{eqnarray}
This K\"ahler potential gives correctly 
the K\"ahler metric found in 
Ref.\cite{IOS1}. 
Again it is the merit of our formulation to obtain 
the K\"ahler potential directly.

\subsection{
Boojums as a Solution of Double-wall Effective Lagrangian
}
As an application of the effective Lagrangian on the domain wall, 
we can consider $1/2$ BPS lump (semi-local vortex) on the 
double wall configuration. 
The resulting configuration is of course a $1/4$ BPS lumps 
stretching between two walls. 
Let us emphasize that we have an effective Lagrangian 
of the double-wall system, rather than an effective 
Lagrangian on a single wall. 
Therefore we can consider lumps stretched between two walls, 
in contrast to the previous studies of lumps on a single 
domain wall such as studies of ``BIon'' \cite{bion}. 
The $1/2$ BPS equation can be derived from the wall 
effective Lagrangian 
with the K\"ahler potential (\ref{eq:kahler_finite}), 
and is given by 
\begin{eqnarray}
\bar \partial \phi_- =0, 
\end{eqnarray}
where $z=x^1+ix^2$ is a complex variable for the two 
spacial dimensions of the lump profile. 
The energy density of the 1/2 BPS state of 
the effective theory is given by 
\begin{eqnarray}
 {\cal E}
={\partial ^2K(\phi ,\phi ^*)\over 
\partial \phi ^i\partial \phi ^{j*}}
\partial _\mu \phi ^i
\partial ^\mu \phi ^{j*}\Big|_{\rm BPS}
=2\partial \bar \partial K(\phi ,\phi ^*)|_{\rm BPS}
={1\over 2}\partial ^2_{(2)}
 K(\phi ,\phi ^*)|_{\rm BPS}, 
\end{eqnarray}
where $\partial ^2_{(2)}$ is the two-dimensional 
Laplacian on the $x^1, x^2$ plane. 
A lump located at the origin $z=0$ 
with the vorticity $k$ and the 
size (and phase) $z_0$ is given by 
\begin{eqnarray}
e^{\frac{\phi _-}2}=(z/z_0)^{k}, 
\quad e^{\phi _+}={\rm const.}. 
\label{eq:lump-moduli-matrix}
\end{eqnarray}
The lump acts as a magnetic charge at $z=0$ on 
the wall, and the flux escapes to infinity on the wall. 
This produces a logarithmic bending of the wall 
and the excitation energy diverges at large $r\equiv|z|$. 
Conversely the logarithmically bent wall can be regarded 
as a lump whose cross section becomes bigger as 
$y \rightarrow \pm \infty$. 
Therefore we introduce a cut-off at the radius 
$r=\Lambda$ to evaluate the energy precisely. 
Defining $ |z_0|=r_0$, 
the energy of the lump inside the radius $\Lambda$ 
is found to be  
\begin{eqnarray}
E_k(\Lambda )
&=&{1\over 2}\int ^{r=\Lambda }_{r=0}dx^2\partial ^2_{(2)}K
=\pi \Big[rK'\Big]^{r=\Lambda }_{r=0}\nn\\
&=&{4\pi c k\over m}\left\{k\log{\Lambda \over r_0}-1
+{\cal O}\left(\left({r_0\over \Lambda }\right)^{2k}\right)
\right\}
\nn\\
&\!\!\!=&\!\!\!
2\pi c\, k L_k\left({\Lambda \over r_0}\right)
-{4\pi \,m\, k\over g^2}
+{\cal O}\left(\left({r_0\over \Lambda }\right)^{2k}\right). 
\end{eqnarray}
This energy diverges in the limit of 
$\Lambda \rightarrow \infty $. 
The divergence gives precisely the energy of 
the semi-local vortex with length 
$\,L_k\left({\Lambda \over r_0}\right)$
$={2k\over m}\log{\Lambda \over r_0}$ 
stretched between two walls, 
since the energy density of the vortex per unit length 
is given by $2\pi c\, k$. 
Moreover, we find a finite contribution 
$-{4\pi \,m\, k/g^2}$. 
This contribution comes from the third term of 
the K\"ahler potential in Eq.(\ref{eq:kahler-density}). 
By using the relation $g^2c=m^2$, the sign and magnitude 
of the contribution is found to agree precisely with 
the contribution from a monopole in the Higgs phase for the 
$U(1)$ case which is called boojum 
\cite{INOS3,Sakai:2005sp,Auzzi:2005yw}. 
Please note that we have obtained the negative contribution 
from the boojum correctly, 
in contrast to the fact that the monopoles 
in non-Abelian gauge theory have positive energy as usual 
\cite{monopoleHiggs}.

\section{Discussion} \label{sc:discussion} 

In the superfield fomalism,  
we have obtained the K\"ahler potential density of 
the effective actions on BPS domain walls (2.60) 
and vortices (3.22) in non-Abelian gauge theory. 
We need the explicit solutions $\Omega$ to 
the master equation (2.46) or (3.19), 
in order to obtain explicit expression of K\"ahler potentials. 
The applicable range is the same with the component formalism. 
However our method have many advantages to the component formalism. 
The most remarkable merit is that the K\"ahler potential can be directly 
obtained with little effort due to manifest supersymmetry. 
This is in contrast to the component formalism, 
in which one has to integrate the K\"ahler metric twice 
to obtain the K\"ahler potential. 
One might consider this is just a technical advantage, but it is not the case. 
It needed tedious calculation to obtain the K\"ahler potentials 
(and even K\"ahler metric), even for the Abelian case [28]. 
Using our method we have been able to obtain the K\"ahler potentials 
very easily even for non-Abelian case, which is a new result. 
Another merit is that we do not have to guess multiplet structure 
of supersymmetry, unlike the case of the component formalism, 
in which identification of fermionic superpartners sometimes 
brings trouble or difficulty.

Extension to composite solitons is an interesting problem. 
For instance loops in domain wall webs \cite{Eto:2005cp} 
or vortices streched between walls \cite{INOS3} 
can have localized modes because they do not change 
boundary conditions.  
Extension to higher derivative terms is also one of 
future directions. 
Higher derivative corrections to translational zero modes 
should sum up to the form of the Nambu-Goto Lagrangian 
if we include infinite number of derivatives 
using nonlinear realizations 
(see for instance \cite{Clark:2002bh}). 
However higher derivative corrections to orientational 
zero modes are not known in general.  
The only exception is a forth order term of 
orientational zero modes of domain walls 
found in a model with degenerate masses \cite{Eto:2005cc}, 
which turns out to be the Skyrme term. 
Relation between the Manton's method discussed in this 
paper with nonlinear realizations 
or mode expansions should be clarified.

\section*{Acknowledgements}

M.~E, M.~N and K.~O would like to thank Nick Manton and 
David Tong for a useful discussion and hospitality at 
DAMTP in Cambridge.  
This work is supported in part by Grant-in-Aid for 
Scientific Research from the Ministry of Education, 
Culture, Sports, Science and Technology, Japan 
No.17540237 (N.S.) 
and 16028203 for the priority area ``origin of mass'' 
(N.S.). 
The work of M.~N. and K.~O. (M.~E. and Y.~I.) is 
supported by Japan Society for the Promotion 
of Science under the Post-doctoral (Pre-doctoral) Research 
Program. 


\appendix
\section{Effective Lagrangian in terms of Component Fields}
\label{sc:component-approach}

Assuming $H^2=0$, we vary the fundamental Lagrangian to 
obtain the equations of motion (with $H^2=0$) as 
\begin{eqnarray}
0&\!\!\!=&\!\!\!
{\cal D}^M{\cal D}_MH^1
+{g^2\over 2}\left(c-H^1H^{1\dagger} \right)H^1
-\Sigma ^2H^1+2\Sigma H^1M-H^1M^2, 
\nn\\
0&\!\!\!=&\!\!\!
{\cal D}^M{\cal D}_M\Sigma 
-{g^2\over 2}\left(\left\{\Sigma ,\,H^1H^{1\dagger} 
\right\}
-2H^1MH^{1\dagger} \right), 
\nn\\
0&\!\!\!=&\!\!\! 
{1\over g^2}{\cal D}^NF_{NM}
-{i\over g^2}[\Sigma ,\,{\cal D}_M\Sigma ]
-{i\over 2}\left(H^1 {\cal D}_MH^{1\dagger} 
-{\cal D}_MH^1 \,H^{1\dagger} \right).
\end{eqnarray}
Using the order assignments (\ref{eq:h1-order}), 
(\ref{eq:h2-order}) and (\ref{eq:field-st-order}), 
we can expand the field equations in powers of $\lambda$. 
The lowest order equations are of order $\lambda ^0$, and 
are automatically satisfied by the BPS equations 
(\ref{BPSeq-H}) and (\ref{BPSeq-Sigma}). 
By taking the order $\lambda$ part of the equations of 
motion, 
we obtain the field equation (\ref{eq:EOM-gauge-field}) 
for the fluctuation $W_\mu$. 
A solution of this equation is given by 
\begin{eqnarray}
W_\mu &\!\!\!=&\!\!\!
i\left((\delta _\mu S^\dagger )
S^{\dagger -1}-S^{-1}(\delta _\mu ^\dagger S)\right), 
\label{eq:sol-gauge-field}
\end{eqnarray}
leading to ${\cal D}_\mu S^{-1}
=-S^{-1}(\delta _\mu \Omega )\Omega ^{-1}$. 
Without the help of the unbroken supersymmetry, it is 
not at all straightforward to find out this soltution, 
contrary to the procedure in 
(\ref{eq:omega-vector})-(\ref{eq:sol-W}) where the 
unbroken supersymmetry has facilitated to obtain the 
solution dramatically. 
Uniqueness of the solution (\ref{eq:sol-gauge-field}) 
comes from the uniqueness of the solution of the master 
equation (\ref{eq:master-eq}). We can confirm that 
the solution (\ref{eq:sol-gauge-field}) leads to the 
following two equations and thus satisfies 
Eq.(\ref{eq:EOM-gauge-field}) 
\begin{eqnarray}
 {\cal D}_\mu H^1
&\!\!\!=&\!\!\!
S^\dagger \delta _\mu \left(\Omega ^{-1}H_0\right)e^{My}, 
\label{eq:cov-der-H}
\\
 {\cal D}_\mu \Sigma +iF_{\mu y}(W)
&\!\!\!=&\!\!\! 
S^\dagger \delta _\mu (\Omega ^{-1}\partial _y\Omega )
S^{\dagger -1}, 
\label{eq:cov-der-Sigma}
\end{eqnarray}
where we use an identity 
\begin{eqnarray}
\Omega ^{-1} \delta _2(\delta _1\Omega \Omega ^{-1})
=\delta _1(\Omega ^{-1}\delta _2\Omega )\Omega ^{-1}, 
\end{eqnarray}
and another relation resulting from the master equation 
(\ref{eq:master-eq}) 
\begin{eqnarray}
\delta \partial _y
\left(\Omega ^{-1}\partial _y\Omega \right)
=-g^2\delta 
\left(\Omega ^{-1}H_0e^{2My}H_0^\dagger \right).
\end{eqnarray} 

The correct gauge transformation of the solution of the 
gauge field is guaranteed by that of $S$: 
$ S\rightarrow S'=SU$ with $U=U(y,\phi (x),\phi ^*(x))$ 
\begin{eqnarray}
 W_\mu \rightarrow W_\mu '
=U^{-1}W_\mu U-iU^{-1}\partial _\mu U, \quad 
U^\dagger =U^{-1}. 
\end{eqnarray}
Let us obtain the effective Lagrangian 
${\cal L}^{\rm eff} \equiv \int dy
{\cal L}
$
by substituting Eqs.(\ref{eq:cov-der-H}) and 
(\ref{eq:cov-der-Sigma}) to the fundamental 
Lagrangian in five-dimensions and by integrating 
over the extra dimension $y$ 
\begin{eqnarray}
&\!\!\! 
&\!\!\! 
{\cal L}^{\rm eff} + T_{\rm w} 
=
\int dy {\rm Tr}
\left[{\cal D^\mu }H{\cal D}_\mu H^\dagger 
+{1\over g^2}
\left({\cal D}^\mu \Sigma -iF^\mu {}_{y}(W)\right)
\left({\cal D}_\mu \Sigma +iF_{\mu y}(W)\right)\right]
\nn\\
&\!\!\!=&\!\!\!\int dy{\rm Tr}\left[\Omega \delta ^\mu
\left(\Omega^{-1}H_0\right)e^{2My}\delta _\mu^\dagger 
\left(H_0^\dagger \Omega ^{-1}\right)
+{1\over g^2}\Omega \delta ^\mu 
\left(\Omega ^{-1}\partial _y\Omega \right)
\Omega ^{-1}\delta _\mu ^\dagger 
\left(\partial _y\Omega \Omega ^{-1}\right)\right]
\nn\\
&\!\!\!=&\!\!\!
\int dy {\rm Re Tr}\left[c\delta ^\mu 
(\Omega ^{-1}\delta _\mu ^\dagger \Omega _0)+ 
{\partial _y^2\over 2g^2}
\left((\delta ^\mu \Omega )\Omega ^{-1}
(\delta _\mu ^\dagger \Omega )\Omega ^{-1}\right)\right]
\nn\\
&\!\!\!=&\!\!\!
\int dy\delta ^\mu \delta _\mu ^\dagger 
\left[(c-{\partial ^2_y\over g^2})\log{\rm det}\Omega 
+ {1\over 2g^2}{\rm Tr}
\left(\Omega ^{-1}\partial _y\Omega \right)^2 \right]
-{1\over g^2}{\rm Re}{\rm Tr}
\left[\Omega ^{-1}\partial _y \Omega \delta ^\mu 
\left(\Omega ^{-1}\delta _\mu ^\dagger 
\Omega \right)\right]\Big|^{\infty }_{-\infty }
\nn\\
&\!\!\!\equiv &\!\!\!
\delta ^\mu \delta _\mu ^\dagger K(\phi ,\phi ^*)=
K_{ij}(\phi ,\phi ^*)\partial ^\mu 
\phi ^i\partial _\mu \phi ^{j*}, 
\label{eq:kahler-metric-wall}
\end{eqnarray}
which gives the K\"ahler potential $K$ as 
\begin{eqnarray}
 K(\phi ,\phi ^*)=\int dy \left[c\log{\rm det}\Omega 
+ {1\over 2g^2}{\rm Tr}
\left(\Omega ^{-1}\partial _y\Omega \right)^2 \right]. 
\label{eq:kahler-potential-wall}
\end{eqnarray}

Let us list some useful formulas. 
The definition of $S$ in Eq.(\ref{eq:comp-gauge-tr}) 
gives 
\begin{eqnarray}
 ({\cal D}_y+\Sigma )S^{-1}=0 \quad 
\rightarrow \quad 
\Sigma ={1\over 2}S^{-1}\partial _y\Omega S^{\dagger -1}. 
\end{eqnarray}
If we consider $[{\cal D}_y,{\cal D}_y+\Sigma ]S^{-1}$, 
we obtain 
\begin{eqnarray}
{\cal D}_y\Sigma 
={1\over 2}S^{-1}\partial _y[(\partial _y\Omega )
\Omega ^{-1}]S .
\end{eqnarray}

\section{Variation of Superfield Form of 6D Lagrangian}
\label{sc:variation-6d-Lag}

In this appendix we summarize useful formulas for the 
variation of vector superfields, especially the variation 
due to gauge transformations. 
Let us first evaluate the exponential of vector superfield 
\begin{eqnarray}
 \delta e^{z V}
&\!\!\!=&\!\!\!
\int _0^1dte^{ztV}z\delta Ve^{z(1-t)V}
=
\left({e^{zL_V}-1\over L_V}\times \delta V\right)e^{zV} 
\nn\\
&\!\!\!=&\!\!\!
e^{zV}\left({1-e^{-zL_V}\over L_V}\times \delta V\right)
=
e^{zV}\left({1\over L_V}\times\delta V\right)
-\left({1\over L_V}\times\delta V\right)e^{zV},
\end{eqnarray}
where the operation $L_V$ is defined in 
Eq.(\ref{eq:adjoint-operation}). 

We need to introduce the Wess-Zumino-Witten like term 
(\ref{eq:WZW-term}) in order to achieve gauge invariance 
under the complexified $U(N_{\rm C})$ gauge transformations. 
Let us denote a term without ${\bf \Phi}$, $c$ or 
${\bf H}^i$ in the $\int d^4\theta$ terms as ${\cal L}_1$ 
\begin{eqnarray}
{\cal L}_1 \equiv 
\int d ^4\theta {\rm Tr}\biggl[
{2\over g^2}
e^{-2{\bf V}}{\bar \partial}e^{2{\bf V}}
e^{-2{\bf V}}{\partial}e^{2{\bf V}}
\biggr]
=
\int d ^4\theta {\rm Tr}\biggl[
{4\over g^2}
\bar\partial {\bf V}
{\cosh(2L_{\bf V})-1\over L^2_{{\bf V}}}\partial {\bf V}
\biggr]. 
\label{eq:Lag-1}
\end{eqnarray}
We can combine this term with the Wess-Zumino-Witten like 
term and obtain 
\begin{eqnarray}
{\cal L}_2 
&\!\!\!
\equiv 
&\!\!\!
{\cal L}_{1}+{\cal L}_{\rm WZW}
=
\int d ^4\theta {\rm Tr}\biggl[
{4\over g^2}
\bar\partial {\bf V}{e^{2L_{{\bf V}}}-1
-2L_{\bf V}\over L^2_{\bf V}}
\partial {\bf V}\biggr]
\nonumber \\
&\!\!\!
=
&\!\!\!
\int d ^4\theta 
\int ^1_0dx\int ^x_0dy
{\rm Tr}\biggl[
{16\over g^2}
\bar\partial {\bf V}e^{2yL_{{\bf V}}}
\partial {\bf V}\biggr]. 
\label{eq:lag-2}
\end{eqnarray}
Thus we obtain the fundamental Lagrangian 
(\ref{eq:fund-Lag-combined}) after combining with the 
Wess-Zumino-Witten like term. 

To obtain the variation of the fundamental Lagrangian 
(\ref{eq:fund-Lag-combined}), it is useful to vary the 
above term (\ref{eq:lag-2}) 
\begin{eqnarray}
\delta {\cal L}_2&\!\!\!=&\!\!\!-\int ^1_0dx\int ^x_0dy
{\rm Tr}\biggl[
{16\over g^2}
 \delta {\bf V}\nn\\
&\!\!\!\times&\!\!\! 
\left(\bar\partial (e^{2yL_{\bf V}}\partial {\bf V})
+\partial (e^{-2yL_{\bf V}}\bar\partial {\bf V})
+{e^{2yL_{\bf V}}-1\over L_{\bf V}}
\left[(e^{-2yL_{\bf V}}\bar\partial {\bf V}),
\,\partial {\bf V}\right]\right)\biggr]. 
\end{eqnarray}
We can use the following formulas to simplify the variation 
\begin{eqnarray}
\int ^x_0dy
\left(\bar\partial (e^{2yL_{\bf V}}\partial {\bf V})
+\partial (e^{-2yL_{\bf V}}\bar\partial {\bf V})\right)
=\bar \partial \left({e^{2xL_{\bf V}}-1\over 2L_{\bf V}}
\partial {\bf V}\right)
+ \partial \left({1-e^{-2xL_{\bf V}}\over 2L_{\bf V}}\bar 
\partial {\bf V}\right) ,
\end{eqnarray}
\begin{eqnarray}
&\!\!\!&\!\!\! \int ^x_0dy
{e^{2yL_{\bf V}}-1\over L_{\bf V}}
\left[(e^{-2yL_{\bf V}}\bar\partial {\bf V}),
\,\partial {\bf V}\right]
=
\int ^x_0dy
{1\over L_{\bf V}}\left\{\left[\bar\partial {\bf V},
\,e^{2yL_{\bf V}}\partial {\bf V}\right]
+\left[\partial {\bf V},\,e^{-2yL_{\bf V}}
\bar\partial {\bf V}\right]\right\} 
\nonumber 
\\ 
&\!\!\!&\!\!\! \quad ={1\over L_{\bf V}}
\left\{\left[\bar\partial {\bf V},
\,{e^{2xL_{\bf V}}-1\over 2L_{\bf V}}\partial {\bf V}\right]
+\left[\partial {\bf V},
\,{1-e^{-2xL_{\bf V}}\over 2L_{\bf V}}
\bar\partial {\bf V}\right]\right\}
\nonumber 
\\
&\!\!\!&\!\!\! \quad =
-{\bar\partial }\left({1\over 2L_{\bf V}}\right)
[\left(e^{2xL_{\bf V}}-1\right)\partial {\bf V}]
-{\partial }\left({1\over 2L_{\bf V}}\right)
[\left(1-e^{-2xL_{\bf V}}\right)\bar\partial {\bf V}] .
\end{eqnarray}
Thus we obtain the variation of the term ${\cal L}_2$ 
\begin{eqnarray}
 \delta {\cal L}_2
&\!\!\!=&\!\!\!
4\delta \int ^1_0dx\int ^x_0dy{\rm Tr}
\left[\bar\partial {\bf V}e^{2yL_{\bf V}}
\partial {\bf V}\right]
\nn\\
&\!\!\!=&\!\!\!
-2\int _0^1dx{\rm Tr}
\left[\delta {\bf V}{1\over L_{\bf V}}
\left\{
\bar \partial 
\left(e^{2xL_{\bf V}}\partial {\bf V}\right)
-\partial \left(e^{-2xL_{\bf V}}
\bar\partial {\bf V}\right)\right\}\right]
\nn\\
&\!\!\!=&\!\!\!
-{\rm Tr}\left[\delta {\bf V}{1\over L_{\bf V}}
\left\{\bar \partial \left((\partial e^{2{\bf V}})
e^{-2{\bf V}}\right)
-\partial \left(e^{-2{\bf V}}\bar\partial 
e^{2{\bf V}}\right)\right\}\right]\nn\\
&\!\!\!=&\!\!\!
-{\rm Tr}\left[\delta {\bf V}{e^{2L_{\bf V}}-1
\over L_{\bf V}}
\partial \left(e^{-2{\bf V}}
\bar\partial e^{2{\bf V}}\right)\right]
=
-{\rm Tr}\left[e^{-2{\bf V}}(\delta e^{2{\bf V}})
\partial \left(e^{-2{\bf V}}\bar\partial 
e^{2{\bf V}}\right)\right] .
\end{eqnarray}

\newcommand{\J}[4]{{\sl #1} {\bf #2} (#3) #4}
\newcommand{\andJ}[3]{{\bf #1} (#2) #3}
\newcommand{\AP}{Ann.\ Phys.\ (N.Y.)}
\newcommand{\MPL}{Mod.\ Phys.\ Lett.}
\newcommand{\NP}{Nucl.\ Phys.}
\newcommand{\PL}{Phys.\ Lett.}
\newcommand{\PR}{ Phys.\ Rev.}
\newcommand{\PRL}{Phys.\ Rev.\ Lett.}
\newcommand{\PTP}{Prog.\ Theor.\ Phys.}
\newcommand{\hep}[1]{{\tt hep-th/{#1}}}


\begin{thebibliography}{10}

  \bibitem{HoravaWitten}     
    P.~Horava and E.~Witten, 
     Nucl.\ Phys.\ {\bf B460}, 506 (1996) [arXiv:hep-th/9510209]. 

  \bibitem{ADD}
    N.~Arkani-Hamed, S.~Dimopoulos and G.~R.~Dvali,
     Phys.\ Lett.\ {\bf B429}, 263 (1998) [arXiv:hep-ph/9803315]; 
    I.~Antoniadis, N.~Arkani-Hamed, S.~Dimopoulos and G.~R.~Dvali,
     Phys.\ Lett.\ {\bf B436}, 257 (1998) [arXiv:hep-ph/9804398].

 \bibitem{RandallSundrum}L.~Randall and R.~Sundrum, 
             Phys. Rev. Lett. {\bf 83}, 3370 (1999) [arXiv:hep-ph/9905221]; 
             Phys. Rev. Lett. {\bf 83}, 4690 (1999) [arXiv:hep-th/9906064].


  \bibitem{DGSW}
    S.~Dimopoulos and H. Georgi, 
     Nucl.\ Phys.\ {\bf B193}, 150 (1981); 
    N.~Sakai, 
     Z.\ f.\ Phys.\ {\bf C11}, 153 (1981);
    E.~Witten, 
     Nucl.\ Phys.\ {\bf B188}, 513 (1981);
    S.~Dimopoulos, S.~Raby and F.~Wilczek, 
     Phys.\ Rev.\ {\bf D24}, 1681 (1981).


\bibitem{Cvetic:1991vp}
  M.~Cvetic, F.~Quevedo and S.~J.~Rey,
  Phys.\ Rev.\ Lett.\  {\bf 67}, 1836 (1991); 
M.~Cvetic, S.~Griffies and S.~J.~Rey,
Nucl.\ Phys.\ B {\bf 381}, 301 (1992) [arXiv:hep-th/9201007].


\bibitem{Dvali:1996bg}
  G.~R.~Dvali and M.~A.~Shifman,
  Nucl.\ Phys.\ B {\bf 504}, 127 (1997)
  [arXiv:hep-th/9611213];
  B.~Chibisov and M.~A.~Shifman,
  Phys.\ Rev.\ D {\bf 56}, 7990 (1997)
  [Erratum-ibid.\ D {\bf 58}, 109901 (1998)]
  [arXiv:hep-th/9706141]; 
  G.~R.~Dvali and M.~A.~Shifman,
  Phys.\ Lett.\ B {\bf 396}, 64 (1997)
  [Erratum-ibid.\ B {\bf 407}, 452 (1997)]
  [arXiv:hep-th/9612128]; 
  A.~Kovner, M.~A.~Shifman and A.~Smilga,
  Phys.\ Rev.\ D {\bf 56}, 7978 (1997)
  [arXiv:hep-th/9706089];
  A.~Smilga and A.~Veselov,
  Phys.\ Rev.\ Lett.\  {\bf 79}, 4529 (1997)
  [arXiv:hep-th/9706217];
  D.~Bazeia, H.~Boschi-Filho and F.~A.~Brito,
  JHEP {\bf 9904}, 028 (1999)
  [arXiv:hep-th/9811084];
  V.~S.~Kaplunovsky, J.~Sonnenschein and S.~Yankielowicz,
  Nucl.\ Phys.\ B {\bf 552}, 209 (1999)
  [arXiv:hep-th/9811195];
  G.~R.~Dvali, G.~Gabadadze and Z.~Kakushadze,
  Nucl.\ Phys.\ B {\bf 562}, 158 (1999)
  [arXiv:hep-th/9901032];
  M.~Naganuma and M.~Nitta,
  Prog.\ Theor.\ Phys.\  {\bf 105}, 501 (2001)
  [arXiv:hep-th/0007184];
  D.~Binosi and T.~ter Veldhuis,
  Phys.\ Rev.\ D {\bf 63}, 085016 (2001)
  [arXiv:hep-th/0011113]; 
  N.~Maru, N.~Sakai, Y.~Sakamura and R.~Sugisaka,
  Nucl.\ Phys.\ B {\bf 616}, 47 (2001)
  [arXiv:hep-th/0107204];
  M.~Naganuma, M.~Nitta and N.~Sakai,
  Phys.\ Rev.\ D {\bf 65}, 045016 (2002)
  [arXiv:hep-th/0108179]; 
  arXiv:hep-th/0210205;
  Y.~Sakamura,
  Nucl.\ Phys.\ B {\bf 656}, 132 (2003)
  [arXiv:hep-th/0207159];
  JHEP {\bf 0304}, 008 (2003)
  [arXiv:hep-th/0302196].







\bibitem{N=1wall-moduli}
  M.~A.~Shifman,
  Phys.\ Rev.\ D {\bf 57}, 1258 (1998)
  [arXiv:hep-th/9708060];
  M.~A.~Shifman and M.~B.~Voloshin,
  Phys.\ Rev.\ D {\bf 57}, 2590 (1998)
  [arXiv:hep-th/9709137];
B.~S.~Acharya and C.~Vafa,
arXiv:hep-th/0103011;
A.~Ritz, M.~Shifman and A.~Vainshtein,
Phys.\ Rev.\ D {\bf 66}, 065015 (2002)
[arXiv:hep-th/0205083];
%
Phys.\ Rev.\ D {\bf 70}, 095003 (2004)
[arXiv:hep-th/0405175];
%
A.~Ritz,
JHEP {\bf 0310}, 021 (2003)
[arXiv:hep-th/0308144].

\bibitem{N1SUGRAwall}
    M.~Eto,  N.~Maru, N.~Sakai and T.~Sakata, 
  Phys.~Lett. {\bf B553}, 87 (2003) 
  [arXiv:hep-th/0208127]; 
    M.~Eto,  N.~Maru, and N.~Sakai, 
 Nucl.Phys. {\bf B673}, 98 (2003), 
  [arXiv:hep-th/0307206]; 
  M.~Eto and N.~Sakai,
  Phys.\ Rev.\ D {\bf 68}, 125001 (2003)
  [arXiv:hep-th/0307276].




\bibitem{N=2walls} 
E.~R.~C.~Abraham and P.~K.~Townsend,
Phys.\ Lett.\ B {\bf 291}, 85 (1992); 
Phys.\ Lett.\ B {\bf 295}, 225 (1992);
  J.~P.~Gauntlett, D.~Tong and P.~K.~Townsend,
  Phys.\ Rev.\ D {\bf 64}, 025010 (2001)
  [arXiv:hep-th/0012178]; 
%
              D.~Tong, 
               JHEP {\bf 0304}, 031 (2003) 
               [arXiv:hep-th/0303151];
               K.~S.~M.~Lee, 
               Phys.\ Rev.\ D {\bf 67}, 045009 (2003) 
               [arXiv:hep-th/0211058];
M.~Shifman and A.~Yung,
Phys.\ Rev.\ D {\bf 70}, 025013 (2004)
[arXiv:hep-th/0312257].

\bibitem{To}  
D.~Tong, 
               Phys.\ Rev.\ D {\bf 66}, 025013 (2002)  
               [arXiv:hep-th/0202012]. 
               
\bibitem{ANNS} 
M.~Arai, M.~Naganuma, M.~Nitta, and N.~Sakai, 
   Nucl.\ Phys.\ B {\bf 652},  35 (2003) [arXiv:hep-th/0211103]; 
``BPS Wall in N=2 SUSY Nonlinear Sigma Model with Eguchi-Hanson Manifold''
in Garden of Quanta - In honor of Hiroshi Ezawa, 
Eds. by J.~Arafune et al. 
(World Scientific Publishing Co. Pte. Ltd. Singapore, 2003) 
pp 299-325, [arXiv:hep-th/0302028]; 
M. Arai, E.~Ivanov and J.~Niederle, 
              Nucl.\ Phys.\ B {\bf 680},  23 (2004) 
              [arXiv:hep-th/0312037]. 
              
\bibitem{IOS1} 
Y.~Isozumi, K.~Ohashi, and N.~Sakai, 
  JHEP\ {\bf 0311}, 060 (2003) 
  [arXiv:hep-th/0310189]. 

\bibitem{IOS2} Y.~Isozumi, K.~Ohashi, and N.~Sakai, 
 JHEP\ {\bf 0311}, 061 (2003)
  [arXiv:hep-th/0310130]. 

\bibitem{INOS1}
  Y.~Isozumi, M.~Nitta, K.~Ohashi and N.~Sakai,
  Phys.~Rev.~Lett.~{\bf 93} 161601 (2004), 
  [arXiv:hep-th/0404198].


\bibitem{INOS2} 
  Y.~Isozumi, M.~Nitta, K.~Ohashi and N.~Sakai,
 Phys.~Rev.~{\bf D70} (2004) 125014, 
  [arXiv:hep-th/0405194].


%
\bibitem{INOS4}
Y.~Isozumi, M.~Nitta, K.~Ohashi and N.~Sakai, 
Proceedings of 12th International Conference on 
Supersymmetry and Unification of Fundamental Interactions 
(SUSY 04), Tsukuba, Japan, 17-23 Jun 2004,  
edited by K. Hagiwara {\it et al.} (KEK, 2004) p.1 - p.16 
 [arXiv:hep-th/0409110]; 
pages 229-238, in ``Themes in Unification'', the Pran Nath Festschrift, 
 (2005) World Scientific, Singapore, 
PASCOS 2004 at Northeastern University, Boston, USA, 
[arXiv:hep-th/0410150]. 

\bibitem{Lambert:1999ix}
  N.~D.~Lambert and D.~Tong,
  Nucl.\ Phys.\ B {\bf 569}, 606 (2000)
  [arXiv:hep-th/9907098].

\bibitem{Eto:2004vy}
  M.~Eto, Y.~Isozumi, M.~Nitta, K.~Ohashi, K.~Ohta and N.~Sakai,
  Phys.\ Rev.\ D {\bf 71}, 125006 (2005)
  [arXiv:hep-th/0412024].

\bibitem{Eto:2005wf}
  M.~Eto, Y.~Isozumi, M.~Nitta, K.~Ohashi, K.~Ohta, N.~Sakai and Y.~Tachikawa,
  Phys.\ Rev.\ D {\bf 71}, 105009 (2005)
  [arXiv:hep-th/0503033].


\bibitem{SakaiYang}
N.~Sakai and Y.~Yang, 
 Comm.\ Math.\ Phys.\ (in press) 
[arXiv:hep-th/0505136].


\bibitem{Hanany:2005bq}
  A.~Hanany and D.~Tong,
  arXiv:hep-th/0507140.

\bibitem{Eto:2005hs}
  M.~Eto, Y.~Isozumi, M.~Nitta, K.~Ohashi and N.~Sakai,
AIP Conf. Proc. 805 (2005) 266-272,
[arXiv:hep-th/0508017].


\bibitem{N2SUGRAwall}
    M.~Arai,  S.~Fujita, M.~Naganuma, and N.~Sakai, 
  Phys.~Lett. {\bf B556}, 192 (2003) 
  [arXiv:hep-th/0212175]; 
    M.~Eto,  S.~Fujita, M.~Naganuma, and N.~Sakai, 
 Phys.Rev. {\bf D69}, 025007 (2004),
  [arXiv:hep-th/0306198]. 





\bibitem{Abrikosov:1956sx}
A.~A.~Abrikosov,
Sov.\ Phys.\ JETP {\bf 5}, 1174 (1957)
[Zh.\ Eksp.\ Teor.\ Fiz.\  {\bf 32}, 1442 (1957)],
[Reprinted in {\em Solitons and Particles}, Eds. C. Rebbi and G. Soliani
(World Scientific, Singapore, 1984), p. 356].

\bibitem{Nielsen:cs}
H.~B.~Nielsen and P.~Olesen,
Nucl.\ Phys.\ {\bf B61}, 45 (1973),
[Reprinted in {\em Solitons and Particles}, Eds. C. Rebbi and G. Soliani
(World Scientific, Singapore, 1984), p. 365].



\bibitem{Taubes:1979tm}
  C.~H.~Taubes,
  Commun.\ Math.\ Phys.\  {\bf 72}, 277 (1980).

\bibitem{Samols:1991ne}
  T.~M.~Samols,
  Commun.\ Math.\ Phys.\  {\bf 145}, 149 (1992).

\bibitem{Chen:2004xu}
  H.~Y.~Chen and N.~S.~Manton,
  J.\ Math.\ Phys.\  {\bf 46}, 052305 (2005)
  [arXiv:hep-th/0407011].

  
\bibitem{Hanany:2003hp}
A.~Hanany and D.~Tong,
JHEP {\bf 0307}, 037 (2003)
[arXiv:hep-th/0306150].

\bibitem{Auzzi:2003fs}
  R.~Auzzi, S.~Bolognesi, J.~Evslin, K.~Konishi and A.~Yung,
  Nucl.\ Phys.\ B {\bf 673}, 187 (2003)
  [arXiv:hep-th/0307287].
  
\bibitem{Eto:2004ii}
  M.~Eto, M.~Nitta and N.~Sakai,
  Nucl.\ Phys.\ B {\bf 701}, 247 (2004)
  [arXiv:hep-th/0405161].

\bibitem{vortices}
  V.~Markov, A.~Marshakov and A.~Yung,
  Nucl.\ Phys.\ B {\bf 709}, 267 (2005)
  [arXiv:hep-th/0408235];
  A.~Gorsky, M.~Shifman and A.~Yung,
  Phys.\ Rev.\ D {\bf 71}, 045010 (2005)
  [arXiv:hep-th/0412082];
  M.~Shifman and A.~Yung,
  Phys.\ Rev.\ D {\bf 72}, 085017 (2005)
  [arXiv:hep-th/0501211];
  R.~Auzzi, M.~Shifman and A.~Yung,
  arXiv:hep-th/0511150;
  A.~Gorsky, M.~Shifman and A.~Yung,
  arXiv:hep-th/0601131.

\bibitem{Hashimoto:2005hi}
  K.~Hashimoto and D.~Tong,
  JCAP {\bf 0509}, 004 (2005)
  [arXiv:hep-th/0506022].

\bibitem{Eto:2005yh}
  M.~Eto, Y.~Isozumi, M.~Nitta, K.~Ohashi and N.~Sakai,
Phys.\ Rev.\ Lett. {\bf 96}, 161601 (2006)[arXiv:hep-th/0511088].

\bibitem{Eto:2006mz}
  M.~Eto, T.~Fujimori, Y.~Isozumi, M.~Nitta, K.~Ohashi, K.~Ohta and N.~Sakai,
 Phys.\ Rev.\ D {\bf 73}, 085008 (2006)[arXiv:hep-th/0601181].




\bibitem{Gauntlett:2000de}
  J.~P.~Gauntlett, R.~Portugues, D.~Tong and P.~K.~Townsend,
  Phys.\ Rev.\ D {\bf 63}, 085002 (2001)
  [arXiv:hep-th/0008221]; 
M.~Shifman and A.~Yung,
Phys.\ Rev.\ D {\bf 67}, 125007 (2003)
[arXiv:hep-th/0212293].

\bibitem{INOS3}
  Y.~Isozumi, M.~Nitta, K.~Ohashi and N.~Sakai,
  Phys.\ Rev.\ D {\bf 71}, 065018 (2005)
  [arXiv:hep-th/0405129].


\bibitem{Sakai:2005sp}
  N.~Sakai and D.~Tong,
  JHEP {\bf 0503}, 019 (2005)
  [arXiv:hep-th/0501207]. 

\bibitem{Auzzi:2005yw}
  R.~Auzzi, M.~Shifman and A.~Yung,
  Phys.\ Rev.\ D {\bf 72}, 025002 (2005)
  [arXiv:hep-th/0504148].

\bibitem{WittenOlive} 
             E.~Witten and D.~Olive, 
              Phys.~Lett. {\bf 78B} (1978) 97.
%
 \bibitem{BPS} E.~Bogomol'nyi, 
               Sov.\ J.\ Nucl.\ Phys.\  {\bf B24} (1976) 449 
               [Yad. Fiz. {\bf 24}, 861 (1976)];
               M.K.~Prasad and C.H.~Sommerfield, 
               Phys.\ Rev.\ Lett.\  {\bf 35} (1975) 760.
%

\bibitem{Atiyah:1978ri}
M.~F.~Atiyah, N.~J.~Hitchin, V.~G.~Drinfeld and Y.~I.~Manin,
Phys.\ Lett.\ A {\bf 65} (1978) 185.

\bibitem{Corrigan:1983sv}
E.~Corrigan and P.~Goddard,
Annals Phys.\  {\bf 154}  (1984) 253.

\bibitem{Nahm:1979yw}
W.~Nahm,
Phys.\ Lett.\ B {\bf 90} (1980) 413;
Phys.\ Lett.\ B {\bf 93} (1980) 42;
``Selfdual Monopoles And Calorons,''
BONN-HE-83-16
{\it Presented at 12th Colloq. on Group Theoretical Methods in Physics, Trieste, Italy, Sep 5-10, 1983}.



\bibitem{Eto:2006pg}
  M.~Eto, Y.~Isozumi, M.~Nitta, K.~Ohashi and N.~Sakai,
  J.\ Phys.\ A {\bf 39}, R315 (2006)
  [arXiv:hep-th/0602170].




\bibitem{Manton:1981mp}
N.~S.~Manton,
Phys.\ Lett.\ B {\bf 110}, 54 (1982).

\bibitem{MantonSutcliffe}
N.~S.~Manton and P.~Sutcliffe, 
``Topological Solitons,'' 
Cambridge University Press (Cabridge, UK), 2004.


\bibitem{MSS} N.~Marcus, A.~Sagnotti and W.~Siegel, 
             Nucl.\ Phys.\  {\bf B224}, 159 (1983). 

\bibitem{PomarolDimopoulos}
 A.~Pomarol and S.~Dimopoulos, 
 Nucl.~Phys.~{\bf B453}, 83 (1995)
[arXiv:hep-ph/9505302]. 


\bibitem{Mirabelli-Peskin}
 E.~A.~Mirabelli and M.~E.~Peskin, 
     Phys.~Rev.~{\bf D58}, 065002 (1998)  
[arXiv:hep-th/9712214]. 


\bibitem{AGW}
N.~Arkani-Hamed, T.~Gregoire and J.~Wacker, 
             JHEP {\bf 0203}, 055 (2002) 
             [arXiv: hep-th/0101233]. 

\bibitem{MartiPomarol}
    D.~Marti and A.~Pomarol, 
 Phys.~Rev.~{\bf D64}, 105025 (2001), 
[arXiv:hep-th/0106256]. 

\bibitem{Hebecker}
A.~Hebecker,
             Nucl.\ Phys.\  {\bf B632}, 101 (2002)  
             [arXiv:hep-ph/0112230]. 



\bibitem{KakimotoSakai}
  K.~Kakimoto and N.~Sakai,
  Phys.\ Rev.\ D {\bf 68}, 065005 (2003)
  [arXiv:hep-th/0306077].



\bibitem{ANS}
  M.~Arai, M.~Nitta and N.~Sakai,
  Prog.\ Theor.\ Phys.\  {\bf 113}, 657 (2005)
  [arXiv:hep-th/0307274]; 
to appear in the Proceedings of the 3rd International Symposium on Quantum Theory and Symmetries (QTS3), September 10-14, 2003, 
[arXiv:hep-th/0401084];
to appear in the Proceedings of the International Conference on ``Symmetry Methods in Physics (SYM-PHYS10)'' held at Yerevan, Armenia, 13-19 Aug. 2003
[arXiv:hep-th/0401102]; 
to appear in the Proceedings of  
SUSY 2003 held at the University of Arizona, Tucson, AZ, June 5-10, 2003
[arXiv:hep-th/0402065].

\bibitem{WessBagger} J.~Wess and J. ~Bagger,
Supersymmetry and Supergravity, Princeton University Press, 
(1992). 

\bibitem{bion}
C.~G.~Callan and J.~M.~Maldacena,
Nucl.\ Phys.\  {\bf B513}, 198 (1998), 
[arXiv:hep-th/9708147]; 
G.~W.~Gibbons,
Nucl.\ Phys.\  {\bf B514}, 603 (1998), 
[arXiv:hep-th/9709027].

\bibitem{monopoleHiggs}
D.~Tong,
Phys.\ Rev.\  {\bf D69}, 065003 (2004)
[arXiv:hep-th/0307302]; 
R.~Auzzi, S.~Bolognesi, J.~Evslin and K.~Konishi,
Nucl.\ Phys.\  {\bf B686}, 119 (2004)
[arXiv:hep-th/0312233];
M.~Shifman and A.~Yung,
Phys.\ Rev.\  {\bf D70}, 045004 (2004)
[arXiv:hep-th/0403149];
  R.~Auzzi, S.~Bolognesi and J.~Evslin,
  JHEP {\bf 0502}, 046 (2005)
  [arXiv:hep-th/0411074];
  L.~Ferretti and K.~Konishi,
  arXiv:hep-th/0602252.

\bibitem{Eto:2004rz}
  M.~Eto, Y.~Isozumi, M.~Nitta, K.~Ohashi and N.~Sakai,
  Phys.\ Rev.\  {\bf D72}, 025011 (2005) 
  [arXiv:hep-th/0412048].

\bibitem{Eto:2005cp}
  M.~Eto, Y.~Isozumi, M.~Nitta, K.~Ohashi and N.~Sakai,
  Phys.\ Rev.\ D {\bf 72}, 085004 (2005)
  [arXiv:hep-th/0506135];
  Phys.\ Lett.\ B {\bf 632}, 384 (2006)
  [arXiv:hep-th/0508241];
  M.~Eto, Y.~Isozumi, M.~Nitta, K.~Ohashi, K.~Ohta and N.~Sakai,
  AIP Conf.\ Proc.\  {\bf 805} (2005) 354
  [arXiv:hep-th/0509127].

\bibitem{Clark:2002bh}
  T.~E.~Clark, M.~Nitta and T.~ter Veldhuis,
  Phys.\ Rev.\ D {\bf 67}, 085026 (2003)
  [arXiv:hep-th/0208184];
  Phys.\ Rev.\ D {\bf 69}, 047701 (2004)
  [arXiv:hep-th/0209142].


\bibitem{Eto:2005cc}
  M.~Eto, M.~Nitta, K.~Ohashi and D.~Tong,
  Phys.\ Rev.\ Lett.\  {\bf 95}, 252003 (2005)
  [arXiv:hep-th/0508130].


\end{thebibliography}
\end{document}